\documentclass[fleqn,usenatbib]{mnras}

\usepackage[normalem]{ulem}
\usepackage{array}
\usepackage{threeparttable}
\usepackage{multirow}
\usepackage{hyperref}
\usepackage{lineno}
\usepackage{graphicx}
\usepackage{xcolor}
\usepackage{pdflscape}
\usepackage[utf8]{inputenc} 
\usepackage[T1]{fontenc}    
\usepackage{caption}
\usepackage{subcaption}
\usepackage{boldline}
\usepackage{booktabs}
\usepackage{url}
\usepackage{amsmath}	
\usepackage{txfonts}
\usepackage{setspace}
\usepackage[normalem]{ulem}

\def \mysep { -- } 

\usepackage[colorinlistoftodos,prependcaption,textsize=tiny]{todonotes}

\title[A strategy for the II Strong Lensing Challenge]{Developing a Victorious Strategy to the Second Strong Gravitational Lensing Data Challenge}
\author[C.R. Bom et al.]{
C. R. Bom,$^{1, 2}$\thanks{E-mail: debom@cbpf.br (CRB)}
B. M. O. Fraga$^{1}$,
L. O. Dias$^{1}$,
P. Schubert$^{1}$\thanks{\textit{in memoriam}},
M.~Blanco Valentin$^{3}$\thanks{At the time of the study, the original affiliation of this author was $^{1}$},
\newauthor
C.~Furlanetto$^{4}$,
M.~Makler$^{1,5}$,
K.~Teles$^{1}$,
M.~Portes de Albuquerque$^{1}$,
R.~Benton Metcalf$^{6,7}$,\\
$^{1}$ Centro Brasileiro de Pesquisas F\'isicas, Rua Dr. Xavier Sigaud 150, CEP 22290-180, Rio de Janeiro, RJ, Brazil\\
$^{2}$Centro Federal de Educa\c{c}\~ao Tecnol\'ogica Celso Suckow da Fonseca, Rodovia M\'ario Covas, lote J2, quadra J, CEP 23810-000,  Itagua\'i, RJ, Brazil\\
$^{3}$ Electrical and Computer Engineering Department, McCormick School, Northwestern University, 633 Clark St, Evanston, IL 60208\\
$^{4}$ Universidade Federal do Rio Grande do Sul, Departamento de Física, CEP 91501-970, Porto Alegre, RS, Brazil\\
$^{5}$ International Center for Advanced Studies \& Instituto de Ciencias Físicas,  ECyT-UNSAM \& CONICET, 25 de Mayo y Francia. C.P.: 1650, San Martín, Buenos Aires, Argentina\\
$^{6}$Dipartimento di Fisica e Astronomia, Università di Bologna, via Gobetti 93/2, I-40129 Bologna, Italy\\
$^{7}$INAF - Osservatorio di Astrofisica e Scienza dello Spazio di Bologna, via Gobetti 93/3, I-40129 Bologna, Italy
}

\begin{document}
\maketitle




\begin{abstract}
Strong Lensing is a powerful probe of the matter distribution in galaxies and clusters and a relevant tool for cosmography. Analyses of strong gravitational lenses with Deep Learning have become a popular approach due to these astronomical objects' rarity and image complexity. 
Next-generation surveys will provide more opportunities to derive science from these objects and an increasing data volume to be analyzed. However, finding strong lenses is challenging, as their number densities are orders of magnitude below those of galaxies.
Therefore, specific Strong Lensing search algorithms are required to discover the highest number of systems possible with high purity and low false alarm rate.
The need for better algorithms has prompted the development of an open community data science competition named Strong Gravitational Lensing Challenge (SGLC).
This work presents the Deep Learning strategies and methodology used to design the highest-scoring algorithm in the II SGLC.
We discuss the approach used for this dataset, the choice for a suitable architecture, particularly the use of a network with two branches to work with images in different resolutions, and its optimization. We also discuss the detectability limit, the lessons learned, and prospects for defining a tailor-made architecture in a survey in contrast to a general one. Finally, we release the models and discuss the best choice to easily adapt the model to a dataset representing a survey with a different instrument.
This work helps to take a step towards efficient, adaptable and accurate analyses of strong lenses with deep learning frameworks.
\end{abstract}




\section{Introduction}
\label{sec:introduction}


The Strong Gravitational Lensing (SL) effect is a phenomenon produced by massive objects along the line-of-sight, typically matter halos in cluster or galaxy scales, that deflects light from sources farther away.
As a result, those systems commonly present magnified and multiple images of sources, which can be highly distorted in the form of rings or arcs. 

Gravitationally Lensed systems can be used as unique probes in many astrophysical and cosmological studies. For instance, the light deflection produces magnified images acting as a ``gravitational telescope'',  enabling the assessment of distant source objects or features that would be beyond the magnitude limit or resolution of a given survey if not lensed \citep[e.g.,][]{2021ApJ...919...20M, 2020ApJ...900..184A,2018ApJ...852L...7E,2011MNRAS.413..643R,2010MNRAS.404.1247J,Marshal2007, 2008ApJ...673...34P}. 
Lensing systems can also be used as non-dynamical probes of the mass distribution of galaxies \citep[e.g.][]{0004-637X-575-1-87,2002MNRAS.337L...6T,2006ApJ...649..599K}, 
and galaxy clusters \citep[e.g.,][]{1989ApJ...337..621K,1998MNRAS.294..734A,2007MNRAS.376..180N,2010AdAst2010E...9Z,2010ApJ...715L.160C,2010ApJ...723.1678C}, providing a relevant observational probe to dark matter~\citep[see, e.g.,][]{2004MPLA...19.1083M}. 

Due to the Cosmological distances involved, strong lensing has also been used to derive cosmological constraints on the cosmic expansion, dark energy, and dark matter \citep[see, e.g.,][]{2010Sci...329..924J,  1998A&A...330....1B,1999A&A...341..653C,2002MNRAS.337L...6T,2001PThPh.106..917Y,2004MPLA...19.1083M,Schwab2010,Enander2013,2016arXiv160203385P}.
In particular, accurate time-delay distance measurements of multiply-imaged lensed QSO systems have enabled precise measurements of the Universe’s cosmic expansion  \citep{2007ApJ...660....1O,2010ApJ...711..201S, Wong:2019kwg}. More recently, time-delay cosmography was also obtained with the strong lensed supernova ``Refsdal'' \citep{2020ApJ...898...87G}. Furthermore, strong lensing can be used to constrain dark matter models \citep{2012Natur.481..341V, hezaveh2014measuring, gilman2018probing, bayer2018observational}, as well as to assess dark matter substructures along the line-of-sight \citep{2018MNRAS.475.5424D, mccully2017quantifying}.

Those systems' many applications and studies motivated an increasing number of searches for strong lensing systems. Earlier searches have been carried out on high-quality space-based data from the Hubble Space Telescope~(HST), for instance, the Hubble Deep Field \citep[HDF;][]{1996ApJ...467L..73H}, the  Great Observatories Origins Deep Survey \citep[GOODS;][]{2004ApJ...600L.155F}, the  HST Medium Deep Survey \citep{1999AJ....117.2010R}, the HST Archive Galaxy-scale Gravitational Lens Survey \citep{HST_galaxyscale_lens} just to name a few.

However, the abundance of data in ground-based experiments, in particular wide-field surveys, fomented the exploration and identification (by visual inspection or automated search on images) of most of the known high-quality strong lensing candidates and, therefore, the majority of confirmed ones.  Many candidates were identified in the Red-Sequence Cluster Survey~\citep[RCS;][]{2003ApJ...593...48G}, in the  Sloan Digital Sky Survey \citep[SDSS;][]{2007Estrada,2009MNRAS.392..104B, 2010ApJ...724L.137K, 2011RAA....11.1185W,2012ApJ...744..156B}, The Canada-France-Hawaii Telescope Legacy Survey 
\citep[CFHTLS;][]{2007A&A...461..813C, 2012More,Maturi2014,RINGFINDER,SPACEWARPSII,ParaficzCFHTLS}, the  Deep Lens Survey \citep[DLS;][]{Kubo2008}, the Dark Energy Survey \citep[DES; e.g.,][]{DESSL1Nord, Diehl_2017}, and the Kilo Degree Survey \citep[KIDS;][]{2017arXiv170207675P}. The future 2020s and 2030s observatories, like the Vera Rubin \citep{ivezic2008lsst}, Euclid \citep{laureijs2011euclid}, and Nancy Grace Roman Space  \citep{green2012wide} Telescopes, are expected to increase the number of strong lensing candidates by a few orders of magnitude than what is currently known \citep[see, for instance, ][]{2015ApJ...811...20C}.

Many of the earlier catalogues of strong lensing systems were found through visual searches only.
Nevertheless, the current large data sets from wide-field surveys triggered the development of fully or human-assisted automated search methods to find and classify \citep{2015mgm..conf.2088D,cheng2020identifying,avestruz2019automated, pca_lensfinder, RINGFINDER}, and more recently inferring the properties of \citep{2017Natur.548..555H,Bom2019,2021MNRAS.505.4362P,2021arXiv211205278L, schuldt2021holismokes} lens candidates.

Most of those techniques are based on image processing and/or Neural networks. In particular, several works have established that both traditional neural networks \citep{2016arXiv160704644B, 2007Estrada} and deep neural networks \citep{2019MNRAS.484.3879P,2019MNRAS.484.5330J,2019MNRAS.482..807P, challenge01,2018MNRAS.473.3895L, 10.1093/mnras/stx1492, 2021arXiv211201541M} can be used to identify lenses from non-lenses, with minimal human intervention. Although there is a certain intuition on how those techniques work, except when the methods explicitly take morphological features or colours rather than the raw image, it is not completely clear which features are used by Deep Learning algorithms in strong lensing identification. Even so, several searches with Deep Learning techniques have successfully found a large number of candidates~\citep[see,  e.g.,][]{2019MNRAS.484.3879P,2019MNRAS.484.5330J,10.1093/mnras/stx1492}. 
The interest in the field led to a community effort to develop the best solutions in the two Strong Gravitational Lensing Challenges~\citep[henceforth SGLC, ][]{challenge01,challenge02}. Both were data challenges where the participants developed different methods to find Strong Lensing from different sets. The first challenge was performed in $100$k simulated images mimicking KIDS-quality survey data, with a $48$ hour time limit. The second challenge used $100$k images with Euclid-like conditions, including $4$ bands, one of them with a different resolution in terms of pixel scale.

In this work, we present the path that ultimately led our team to develop the winning solution of the second Strong Gravitational Lensing Challenge. We discuss the lessons learned and how to work on the data. Most of our pipeline is generic enough and thus valuable for other potential applications in Deep Learning image processing problems in astronomy. We present how to preprocess the data, what is the deep learning architecture choice, including how to combine images with resolutions in the same network. Additionally, we use a technique to identify what region on the image the algorithm is using to make a classification and therefore, we may assess the decision-making process. We further discuss the adaptability of our model to other datasets and the detection limits in terms of lensed pixels above the background. We make our pipeline and model weights publicly available\footnote{\url{https://github.com/cdebom/cast_lensfinder}}.

This paper is organized as follows: 
in \S\ref{sec:dataexp}, we briefly describe the data and present the initial processing and data exploration. Later, in \S\ref{sec:dlmodels} we introduce the deep learning models, the architecture definition and choice of parameters used in this work.
Then, in \S\ref{sec:training}, we describe the training process, convergence and overfitting. Following that, in \S\ref{sec:performance}, we describe the model's performance in the test set. In \S\ref{sec:lime}, we use a technique to infer which features are relevant to the Deep Learning classification. In \S\ref{sec:limits}, we evaluate the sensitivity of the current method and define the detection limits. Later, in \S\ref{sec:adaptability}, we present a prescription on how to adapt the method for a different dataset. Finally, in \S\ref{sec:conclusions}, we make a discussion and present the conclusions of this work and an outlook for future.

\section{Data Exploration}
\label{sec:dataexp}
\subsection{Catalogue and available data}

The data for the second Strong Gravitational Lensing Data Challenge (II SGLC) is comprised of $100,000$ simulated objects in four different bands: VIS (visible) and H, J, and Y (infrared), for a total of $400,000$ images. The VIS images have a $200\times200$ pixel resolution with a pixel scale of $0.1$ arcsec, while the other bands have $\sim3$ times lower resolution, with $66\times66$ pixel images with $0.3$ arcsec of pixel scale. The images are an update from the I SGLC  \citep[for a detailed description of the datasets, see][]{challenge01,challenge02}.
For convenience, we highlight the main features of the simulation process.
The simulations used dark matter halo catalogues derived from the Millennium Observatory project \citep{millennium_obs} which included subhalos of larger halos. The sources redshifts in the II SGLC were between $1.27$ and $11.08$, with a mean of $2.76$. The catalogues are used as input to the GLAMER lensing code \citep{2014MNRAS.445.1942M,petkova2014glamer}, and the sources are set using Hubble Ultra Deep Field (UDF) images. The lens light profile is defined by semi-analytic models that use the galaxy parameters from The Millennium Observatory simulations, which were set to match Euclid’s expected observational conditions, scales, and pixelization.
During the competition’s time, the organizers provided no details on how the simulations were generated. 
\par Together with the images, a catalogue of properties was also provided, instead of a simple truth table. The catalogue contains, among others, the coordinates of the center of the critical curves, the redshift of source and lens objects, source effective magnification, and number of source pixels with intensity $1\sigma$ above the background level. Thus, we may have  simulated images where just one pixel from the lensed source was visible. For the challenge, a detectable strong lens system was defined by the following criteria:

\begin{itemize}
    \item[--] Number of groups of source larger than zero;
    \item[--] source's effective magnification in all bands larger than $1.6$;
    \item[--] number of source pixels with intensity above $1\sigma$ of the background level larger than $20$.
\end{itemize}
$49213$ images in the dataset followed these criteria and were therefore classified as a lens, making it reasonably balanced. Unlike a simple truth table, a catalogue like this allows us to explore how changing the detectable strong lens definition (e.g., the number of source pixels above the background level) can affect the classification. 
\par
\begin{figure*}
\centering
\includegraphics[width=.9\textwidth]{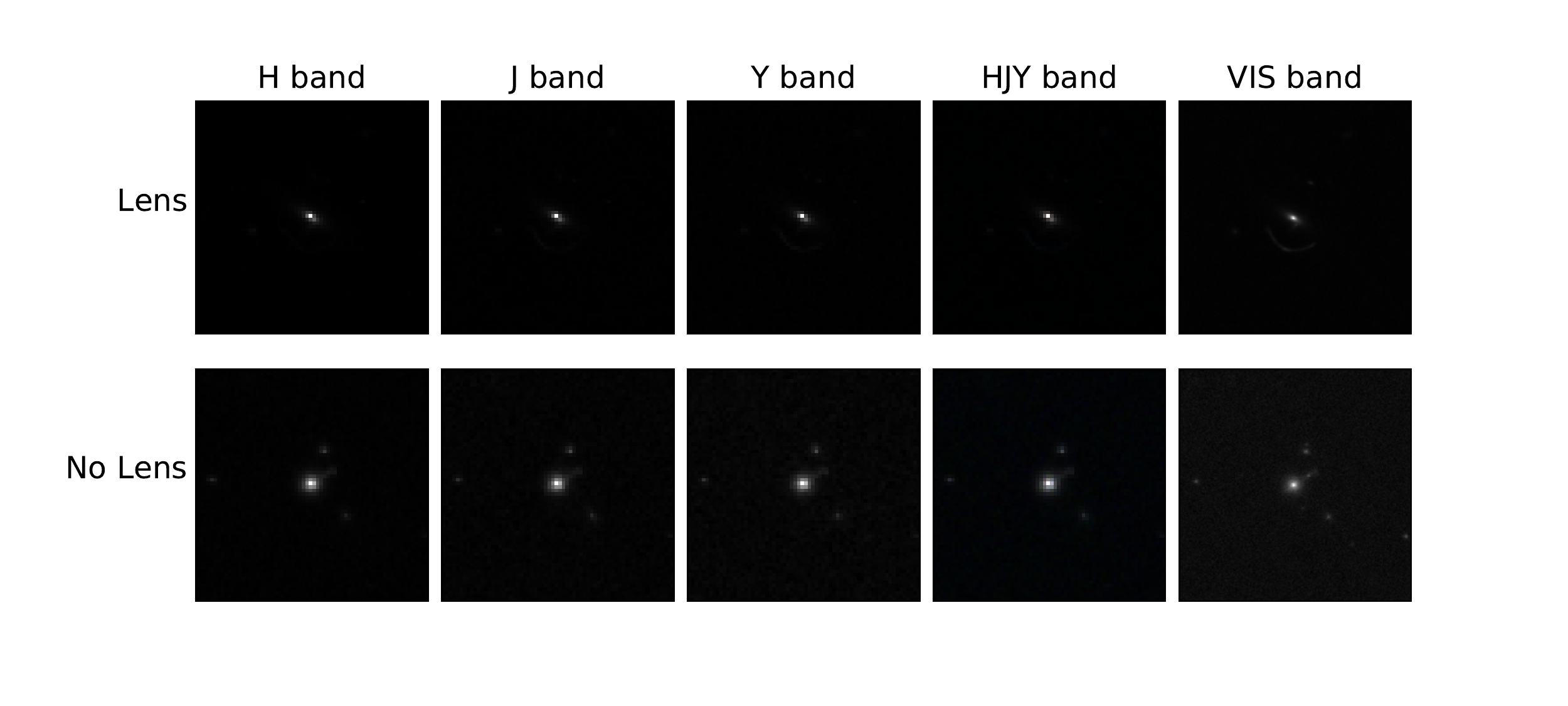}
\caption[Raw H, J, Y, VIS and HJY band images from two objects on dataset, one lens and one with no lens.]{H, J, Y, VIS and HJY band images combined from two objects on dataset.}
\label{fig:img_nopre}
\end{figure*}

 Figure \ref{fig:img_nopre} shows two example objects from the catalogue, one a lens (top row)  and a non-lens object (bottom row), in each of the four bands, and joining H, J and Y together. The images show almost no visible features, and the lensed source is only barely visible to the naked eye in the VIS band as an arc below the galaxy.

\begin{figure}
\centering
\includegraphics[width=.45\textwidth]{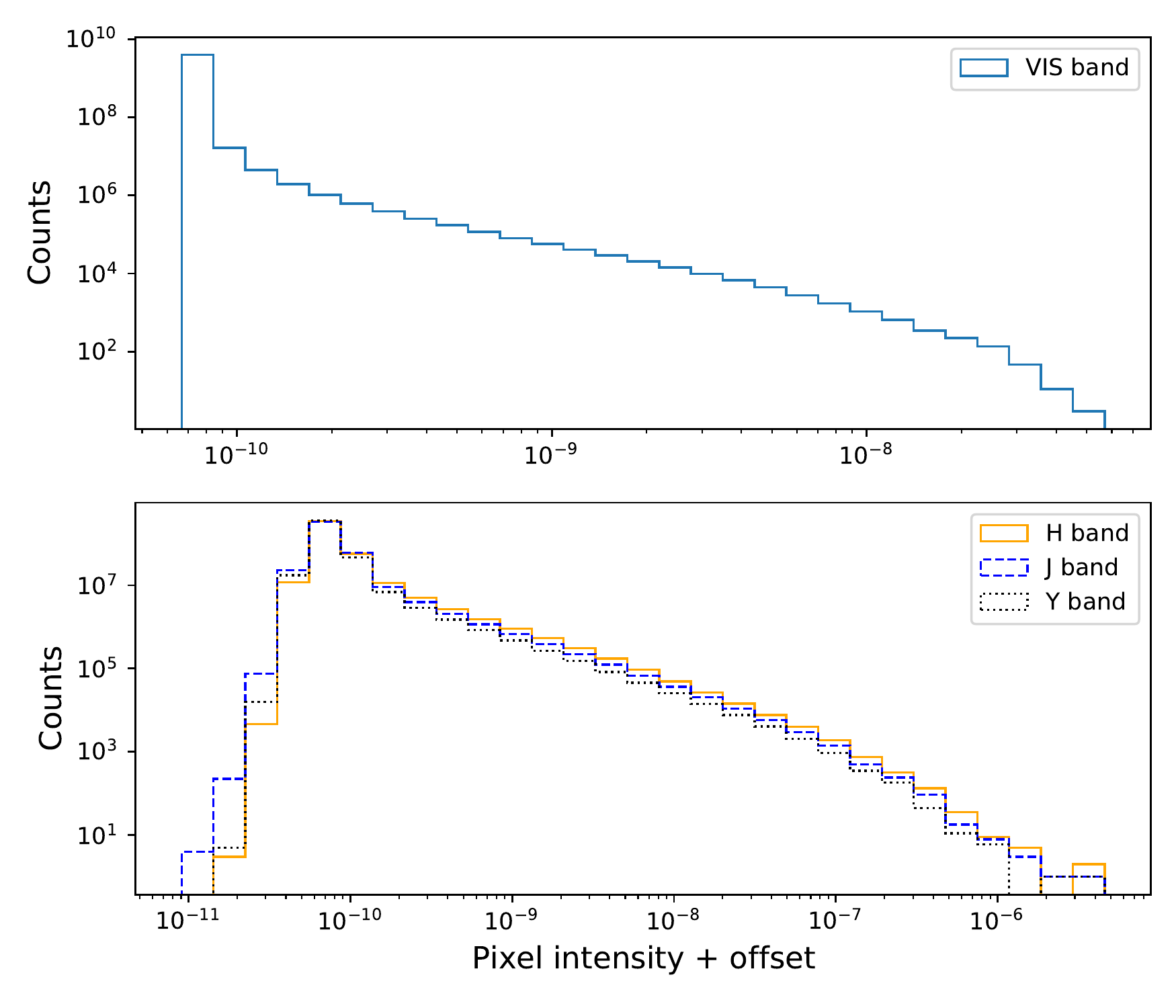}
\caption[Pixel histogram from H, J, Y and VIS image bands]{\small Pixel histogram from H, J, Y and VIS  image bands. The pixel values are offset by a fixed amount to remove negative values and allow the use of a logarithmic scale. }
\label{b_hist}
\end{figure}

The pixel intensity distribution for all four bands of all images is shown in Figure \ref{b_hist}. They are all extremely skewed toward higher values, with the minimum and the median pixel values comparable. At the same time, the maximum is three to four orders of magnitude (depending on the band) larger than the median. Hence, to present the distribution, we use a logarithmic scale so that features are more easily seen. We also apply a constant offset to make all pixel values positive. These extremely skewed distributions help explain the lack of visual features when displaying the images: the brightest pixels are orders of magnitude brighter than the other ones, almost completely dominating the entire image and making the lensing features difficult to identify by visual inspection without any image processing. Furthermore, this extreme difference could make a neural network look only to the high-intensity pixels, almost completely disregarding the rest.

\subsection{Preprocessing}

\par
\begin{figure}
\centering
\includegraphics[width=.45\textwidth]{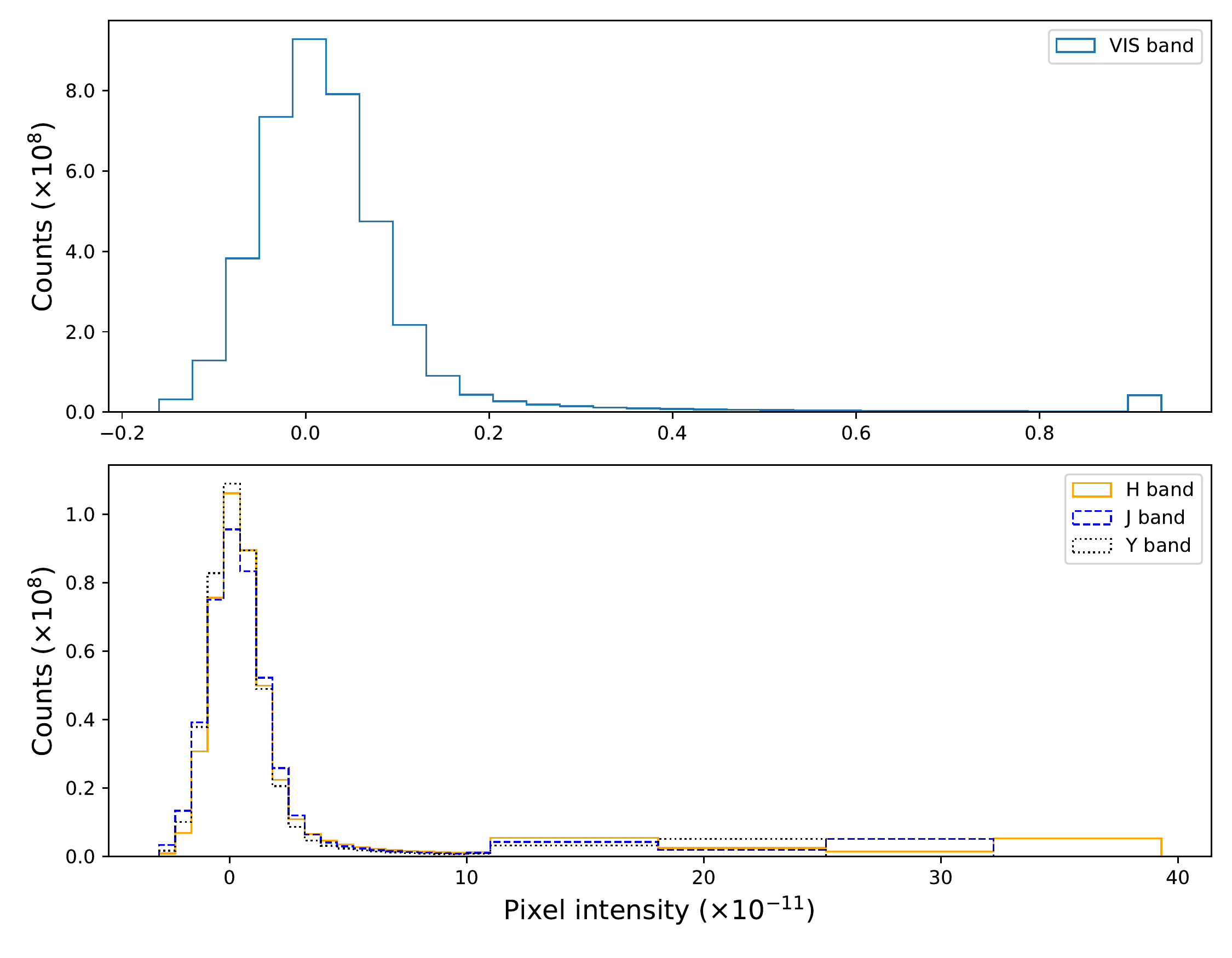}
\caption[Pixel histogram from H, J and Y image bands]{Pixel intensity histogram for VIS, H, J and Y image bands with clipped images.}
\label{fig:c_hist}
\end{figure}

 The deep learning methods are sensitive to visual features by construction, so we preprocess the data to enhance the visualization. For completeness, we also evaluate the performance of our Deep Learning pipeline without any preprocessing. A simple normalization would not change the skewness of the pixel intensity distribution. Therefore, we first make a contrast adjustment by clipping the image histograms, i.e., we choose a lower and an upper bound and set every pixel above or below those values to be equal to the nearest bound. We define two sets of contrast adjustments. The first one was used for the II SGLC: 
since the images present negative pixel values, we chose the lower bound by first inverting the image, taking the 99.9 percentile, and multiplying it by $-1$; the upper bound was set as the $98$ percentile, both using the full pixel intensity distribution of each band. However, upon further visual inspection, we found that the H, J, and Y images were saturated in several cases, making the small features of objects or lenses disappear. After a search around the bounds used in the II SGLC and a visual inspection of some example images after each clipping, we chose a second contrast adjustment to the 0.1 and 99 percentile as the lower and upper bound, respectively. Figure \ref{fig:c_hist} shows the pixel intensity distribution after this clipping; the extremeness of the original distribution can now be appreciated: the 99 percentile for all bands is three to four orders of magnitude lower than the maximum. With the clipping, now the median the maximum are comparable.
\par
\begin{figure*}
\centering
\includegraphics[width=\textwidth]{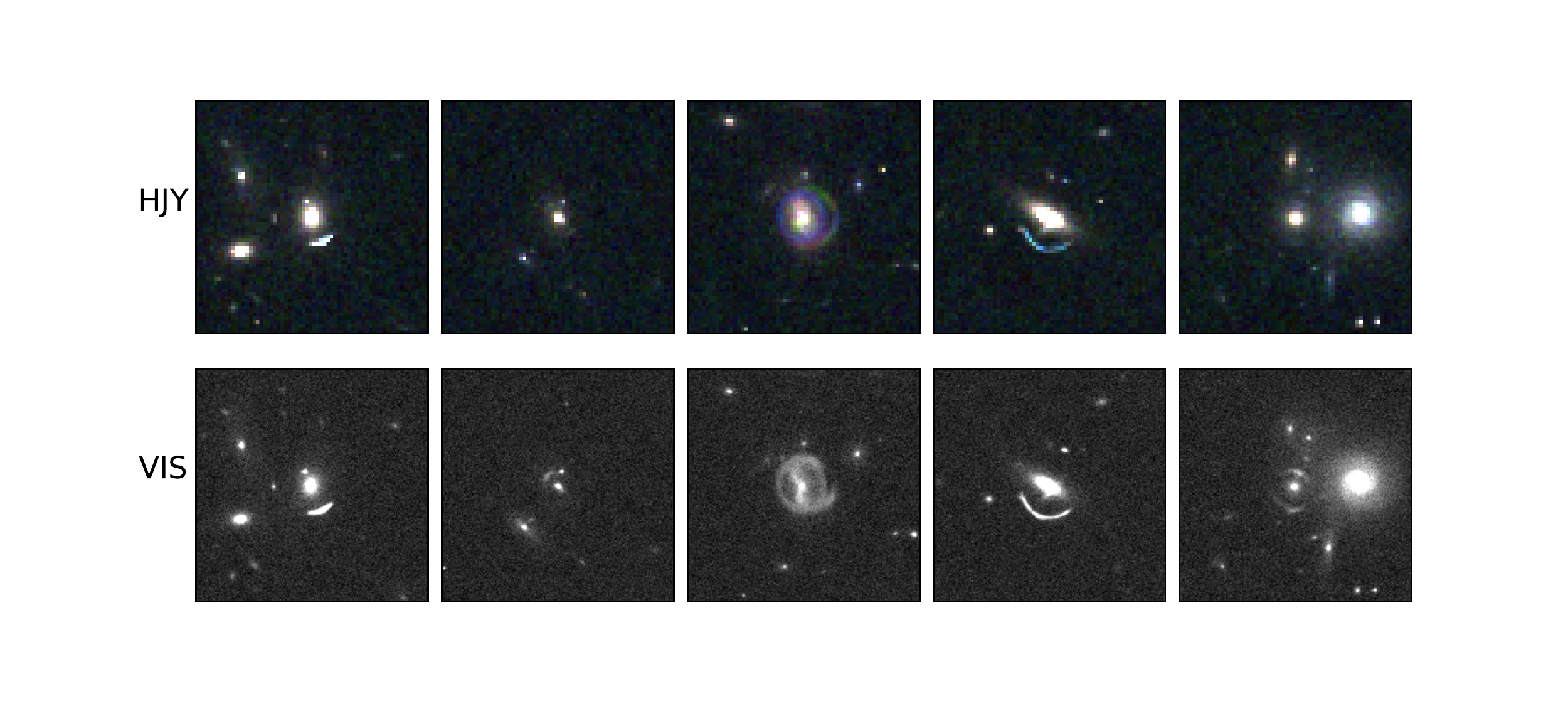}
\caption[H, J, Y, VIS and HJY band images combined from two objects on dataset after pre-processing.]{\small H, J, Y, VIS and HJY band images combined from two objects on dataset after pre-processing.}
\label{fig:img_pre}
\end{figure*}

 The effect in the images can be seen in Figure \ref{fig:img_pre}, which shows five examples of clipped lensed images in the VIS band and combined HJY band, where the fourth column corresponds to the same lens shown in Figure \ref{fig:img_nopre}. Several features are now visible, such as the lenses and other galaxies in the background. However, while some of the lenses are easily noticeable in both VIS and infrared bands, some can only be readily seen in the VIS band, namely the ones in the second and last columns. This suggests that, at least in some cases, resolution might be more important than colour when visually searching for gravitational lenses. While this new preprocessing helped the lensing features visualization and thus motivates its usage, there is no guarantee that it will improve the results before the final DL performance evaluation.

\section{Deep Learning Models}
\label{sec:dlmodels}

\subsection{EfficientNets}
We made use of a family of CNN models known as EfficientNet \citep{efficientnet}, which were built to be high-performing, i.e., state-of-the-art, in benchmarking image classification datasets such as ImageNet, while also being easily scalable. In order to improve a CNN's performance, one can scale up the net in three ways: increasing the number of layers (depth), the number of channels (width), or the input image resolution. Increasing the depth could make the network learn more of the complex features but also makes it more prone to suffer from the vanishing gradients problem, where the derivative of the loss with respect to the weights approach zero, and the learning stalls~\citep{Goodfellow-et-al-2016}. Furthermore, deeper networks tend to saturate on accuracy \citep{he2016deep}. On the other hand, while wide and shallow networks would escape this problem, they would fail to learn more complex features. Therefore, \citet{efficientnet} introduced the idea of Compound Scaling, where depth, width, and image resolution are scaled up at the same time while maintaining a balance between them:
\begin{equation} \label{eq:scaling} 
\begin{aligned}
\text{depth: } & d  = \alpha ^ \phi \\
\text{width: } & w = \beta ^ \phi \\
\text{resolution: } & r   =  \gamma ^ \phi,  \\
\end{aligned}
\end{equation}
where $\phi$ is an integer named compound coefficient and $\alpha$, $\beta$ and $\gamma$ are constants. Floating-point Operations Per Second (FLOPS) on a convolutional operation scale as $d \cdot w^2 \cdot r^2$; thus, for it to scale as $2^{\phi}$ for any new compound coefficient, $\alpha$, $\beta$ and $\gamma$ are subject to the following constraints:

\begin{equation} \label{eq:optobj} 
\begin{aligned}
& \alpha \cdot \beta ^2 \cdot \gamma ^ 2 \approx 2 \\
& \alpha \ge 1, \beta \ge 1, \gamma \ge 1, 
\end{aligned}
\end{equation}

\par Before scaling up, a good base model is needed; \citet{efficientnet} use a multi-objective architecture search, optimizing for accuracy and FLOPS, using the same parameter space as \citet[MnasNet;][]{tan2019mnasnet}. The resulting network is similar to MnasNet but with more parameters, called EfficientNet-B0, having a Mobile Inverted Bottleneck with a squeeze and excitation connections as its building block (see Figure \ref{fig:efnb2}, top right). A small grid search is done to obtain the best values for $\alpha$, $\beta$, and $\gamma$ for B0, which are then left constant. A family of EfficientNets (B1 to B7 originally) can be then built by increasing the compound coefficient $\phi$. EfficientNet-B7 achieved the best results for top-1 accuracy in Imagenet, outperforming more complex architectures in terms of the number of parameters. However, among the EfficientNet model family, many of them already have a high performance, around $80\%$, in top-1 accuracy, i.e. the accuracy in multi-class problems considering as a correct prediction only the highest probability class.
\subsection{Proposed Architecture}
\begin{figure*}
    \centering
    \includegraphics[width=0.27\linewidth]{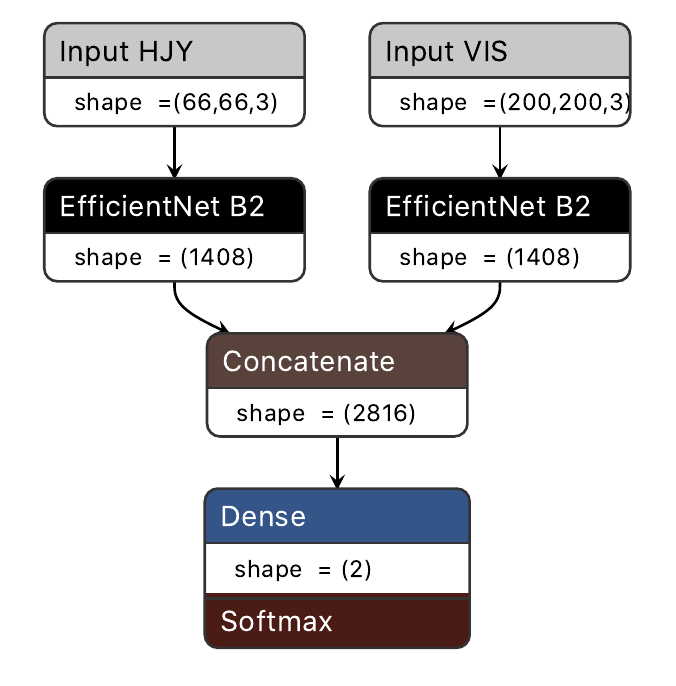}
    \includegraphics[width=0.2\linewidth, angle=90]{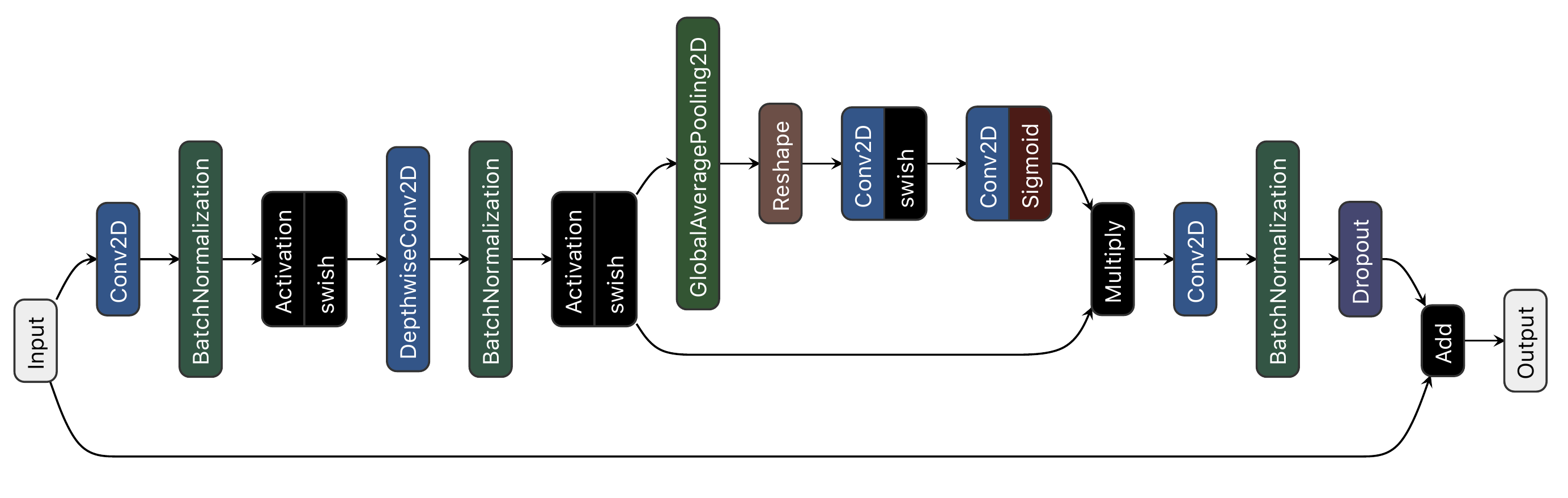}\hfill
    \includegraphics[width=0.2\linewidth, angle=90]{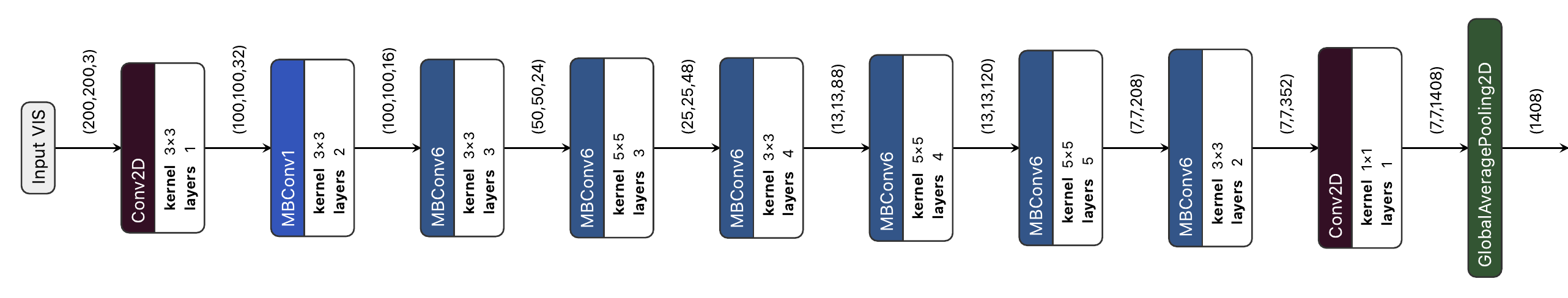}\hfill

    \caption{The Deep Learning architecture used. The model was based on the EfficientNet B2 model. Top left: Our two-branched model used to train both imagens with HJY and VIS bands at the same time. Top Right: Mobile Inverted Bottleneck block, with squeeze and excitation phases. Bottom: Full architecture of the EfficientNet B2 model. MBConv blocks are the Mobile Inverted Bottleneck mentioned previously.}
    \label{fig:efnb2}
\end{figure*}

EfficientNet models defined by \citet{efficientnet} are built to have as inputs three channels. Since H, J, and Y images have the same resolution, it is natural to combine 
them. However, the same cannot be done for the VIS band. Therefore, we define a second EfficientNet model to operate with the VIS images. We use two simple methods to create two extra channels in the VIS model:
\begin{itemize}
    \item Propagate the images to the other channels, effectively creating three equal ones, which we call VIS (repeated);
    \item Fill the other channels with null arrays, which we call VIS (zeros).
\end{itemize}
While this approach is suitable if we want to train either the HJY or VIS bands alone, it does not help if we want to use them together. In order not to scale up or down the images, keeping them in their original form, we chose to create a net with two EfficientNet branches, one for each input. The final model has the last two dense layers of the original architecture removed and concatenates both outputs before passing them through a fully connected layer with softmax activation (see Figure \ref{fig:efnb2} top left). We made an initial test of performance with the EficientNet models and did not find any significant improvement of performance beyond the B2 model, considering cross-validation uncertainties,  while the more complex B3-B7 models have more parameters and can take considerably longer to train. Thus, we use in our branches the EfficientNet-B2 architecture for our main results, which ended up being the model that achieved the highest score on the II SGLC.
Furthermore, it is worth noticing that EfficientNet-B2 was also successfully implemented in image classifications of astrophysical sources for galaxy morphology catalogues as described in \citet{2021MNRAS.507.1937B}. 
\citet{2022MNRAS.509.3966W} used an EfficientNet-B0 model for the same kind of problem.

Hence, we test the following architectures: one-branched EfficientNet, using either VIS or HJY, and two-branched with all bands. We tested the two versions for all cases with VIS band, VIS (repeated) or VIS (zeros). After some visual assessment in the near infrared 
bands we found that many of the objects were visually very similar, so we also added a simpler model where we chose one of the near-infrared band, Y, and VIS, filling the other two channels with null arrays. Therefore, we evaluate a total of six models presented in table \ref{table:results}, each of them with the two preprocessings schemes described in the previous sections and without any preprocessing.

\section{Results}


\subsection{Training}
\label{sec:training}

The previously described models were trained using one of the state-of-art optimizers: the Rectified Adam \citep[RADAM;][]{radam}, with a binary-cross entropy loss. We initialize the networks with pre-trained weights from the ImageNet dataset in order to improve the computing time needed for convergence, the overall performance, and to make the training more stable as in \citet{2021MNRAS.507.1937B}. All images were normalized in the $[0,1]$ range before being fed to the net. A 10-fold cross-validation was performed: the sample is divided in 10 parts; at each iteration, one of these parts is used as a validation sample, while the rest is used for training. This helps to avoid biases that could arise from the selection of specific train/validation sets. For each fold, we trained the net for $50$ epochs with a batch size of $64$. 

Furthermore, we employ data augmentation strategies consisting of random rotations, mirroring in both axes, and zooming in or out the images between $0.8$ and $1.2$ times. We trained the models in a Multi-GPU server with 8 RTX 3090 with 24 GB of GPU memory each. Since cross-validation is a high resource-consuming procedure, we selected a random sample of $20,000$ images (4 bands each) from the full dataset for training/validation. The models and training were implemented in Tensorflow 2 \citep{tensorflow}. The remaining $80$k images are used as an independent test set for network performance assessment.

The training curves with no preprocessing show no or a negligible decrease in the training and validation losses, with the latter having large oscillations in some epochs indicating an underfit. The one branch models containing HJY or VIS have the training loss decrease rapidly in the first few epochs, and more slowly later on; however, there is a strong overfit (validation loss much larger than training loss) already in the first $3-4$ epochs, with the validation loss increasing and having again large oscillations. Training a two-branched network with HJY+VIS or Y+VIS  simultaneously removes this large oscillation, and both losses decrease at approximately the same pace for $15-20$ epochs when we start to see some overfitting.
\par
\begin{figure}
    \centering
    \includegraphics[width=\linewidth]{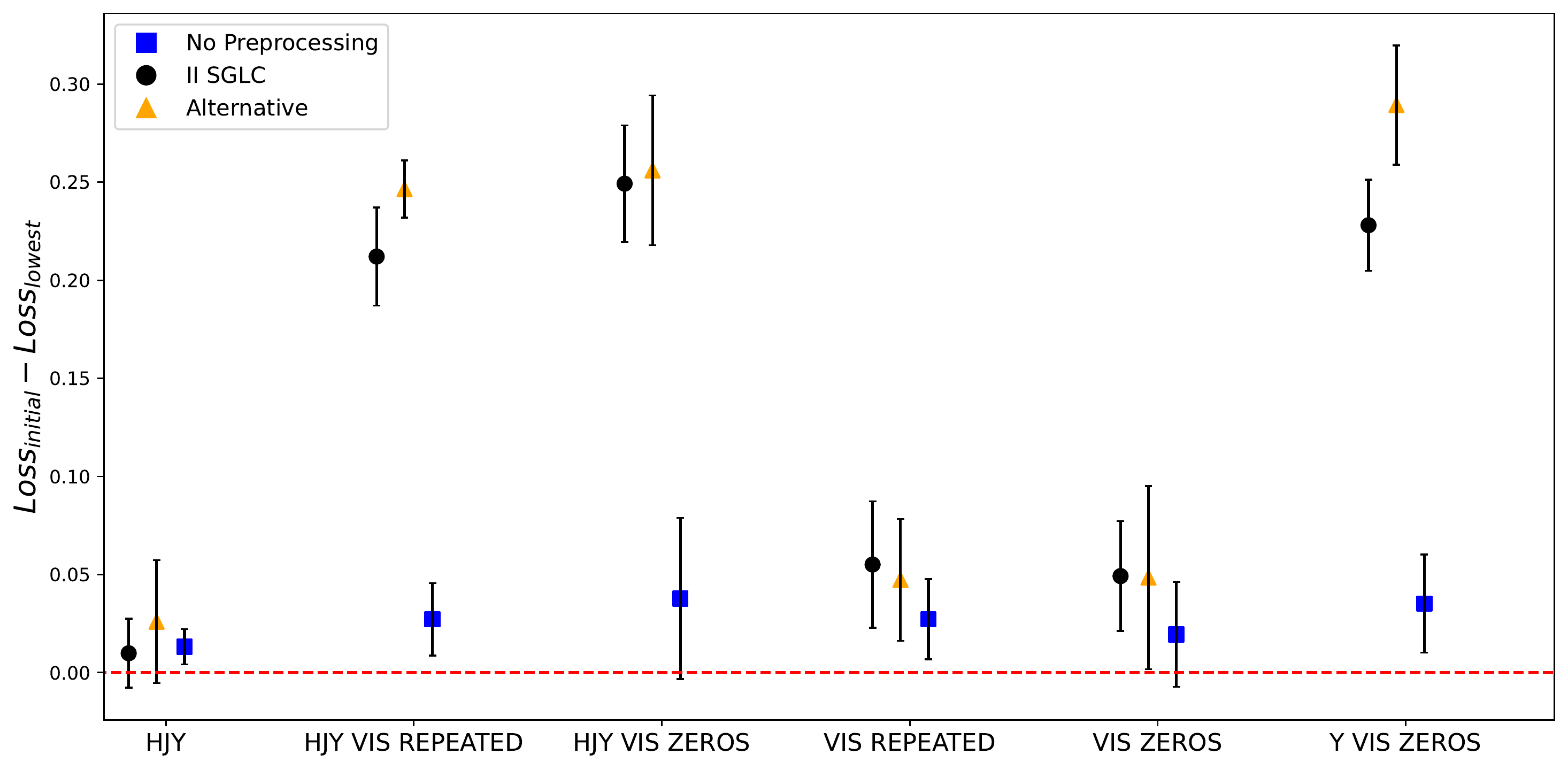}
    \caption{Mean validation loss difference between the initial and the best epoch colour coded for the different preprocessing methods used. Blue: no preprocessing, Black: preprocessing as submitted to the challenge, Orange: new preprocessing. Larger differences indicate that the model learned more.}
    \label{fig:diff_val_loss}
\end{figure}

 To avoid being contaminated by overfitting at each fold, we use the model with the lowest validation loss to make our predictions on the test set. In Figure \ref{fig:diff_val_loss} we present the difference between the initial and the lowest validation loss for each combination of bands and preprocessing. The error bars are standard deviations, as measured in all 10 trainings for each configuration. This quantity is related to how well the model learned the problem and its ability to generalize. Smaller values mean that the network did not learn a more general solution, as the validation loss did not decrease. This happens in all models with no preprocessing, and in models with only one branch (HJY or VIS). Furthermore, the II SGLC preprocessing and the current alternative one presents similar results considering the errors.

\par
\begin{figure}
    \centering
    \includegraphics[width=\linewidth]{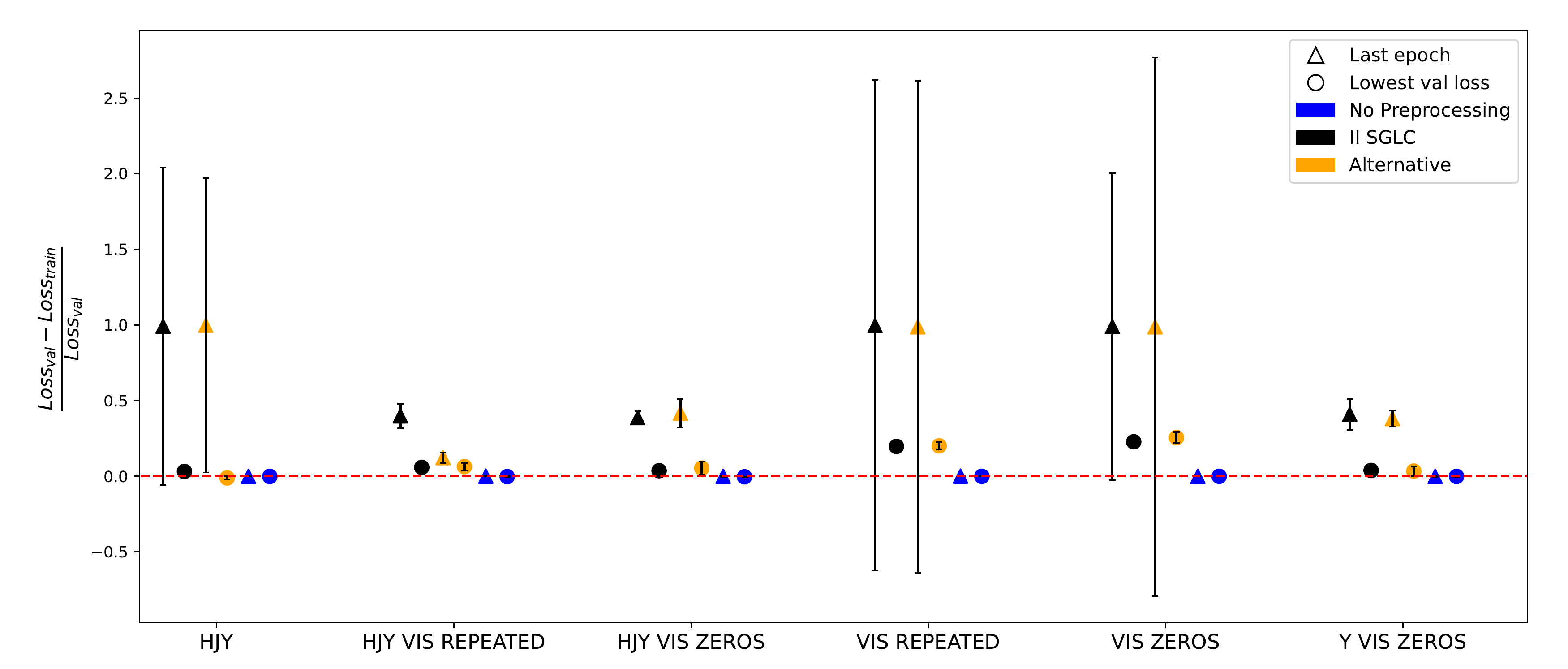}
    \caption{Difference between training and validation mean losses at the last and best epoch, colour coded for the different preprocessing methods used. Blue: no preprocessing, Black: preprocessing as submitted to the challenge, Orange: new preprocessing. Triangles show the difference in the last epoch while circles show it at epoch with the lowest validation loss. Lower differences indicate less overfitting or no learning at all.}
    \label{fig:diff_loss}
\end{figure}

 Overfitting occurs when the difference between training and validation loss starts to diverge. In Figure \ref{fig:diff_loss} we present the difference between the mean validation loss and the mean training loss at the last epoch and at the epoch with the lowest validation loss for every combination of bands and preprocessing. Smaller values suggest that overfitting did not occur or it is negligible. Nevertheless, the overfitting results should also consider the underfitting results from Figure \ref{fig:diff_val_loss}. In a situation of underfitting,  one might also expect that the losses stay nearly constant for all epochs and thus small differences between training and validation loss. Confirming what was seen before, the models with no preprocessing did not learn, as the training and validation loss changed negligibly during training. Furthermore, one-branched models have a volatile training phase, with high variance in the loss at the last epoch and a little overfitting even in the best epoch. Training a two-branched network presents, as before, the best results, with almost no variance between the losses at each fold and little to no overfitting when considering the best epoch. However, it can be seen that for most models, there is a tendency to overfit at later epochs. Again, the II SGLC and the alternative preprocessing are compatible within the errors except in the case of HJY and VIS (repeated), where training with the alternative preprocessing show negligible overfit at all even in the last epoch.

\subsection{Performance evaluation}
\label{sec:performance}
To assess the models' performance we consider several metrics, such as precision, recall, and the false alarm rate. 
The precision can be defined as:
\begin{equation}
\text{P}=\frac{|\{\text{SL}\}\cap\{\text{systems classified as SL}\}|}{|\{\text{systems classified as SL}\}|}.
\end{equation}
The recall can be defined as:
\begin{equation}
\text{R}=\frac{|\{\text{SL}\}\cap\{\text{systems classified as SL}\}|}{|\{\text{SL}\}|}.
\end{equation}

\noindent The false alarm rate is:
\begin{equation}
\text{F}=\frac{|\{\text{Not SL}\}\cap\{\text{systems classified as SL}\}|}{|\{\text{Not SL}\}|}.
\end{equation}

\noindent The Precision is a measurement of how pure the sample is, i.e., the percentage of the elements classified as a given class that are correct. The recall represents how complete is the classified sample or how many elements of a class were correctly identified. The false alarm rate is the percentage of fake detections. The output of the NN pipeline is a number associated with the probability of a given object being a strong lensing system. Therefore, to obtain the aforementioned metrics is then necessary to define a probability threshold $t$: varying this threshold in the range $[0,1]$, one can obtain a curve of P and R and other for R and F, with the latter known as Receiver Operating Characteristic (ROC) curve. This is a typical process to assess the quality of a given classification algorithm \citep[see, e.g.][]{magro2021,cheng2020identifying,challenge01, 2021MNRAS.507.1937B,fraga2021}, and can also be used to define the best threshold balancing the Precision, recall or false alarm rate. Additionally, the Area Under the Curve of the ROC (AUC) is also an intuitive quantity to evaluate the classification performance: a perfect classifier would have ${\rm AUC}=1$. In contrast, for a random choice classifier, we would expect ${\rm AUC}=0.5$. Analogously, the area under the P and R curve can also be used as a quality metric.

We summarize our results in Table \ref{table:results}, obtained in a validation sample at each iteration of the cross-validation. There we show the mean area under the curve (AUC) for the Receiver Operating Characteristic (ROC) and Precision-Recall (PR) curves with the error corresponding to one standard deviation when considering all trainings. We also show the mean and error for the $F_{\beta}$ score, defined as
\begin{equation}
    F_{\beta} = (1+\beta^2) \frac{P\times R}{\beta^2P + R}.
\end{equation}
A lower $\beta$, $\sim 0$ will favor the precision P, while Recall R dominates higher values $>1$, and $\beta=1$ represents the harmonic mean between Precision and Recall. For ranking purposes in the II SGLC, $\beta^2=0.001$.
Since they both depend on the probability threshold $t$ chosen to separate between the classes, we take the maximum value of $F_{\beta}$ at each fold,
\begin{equation}
    F_{\beta} = \underset{t}{\text{max}}\,F_{\beta}(t).
\end{equation}

\par The results in the table \ref{table:results} confirm that the net is unable to learn from inputs without any preprocessing, and it is only slightly better using the HJY bands alone. The results improve when using only the VIS band, with some slightly numerical advantage for VIS with zeros. The comparison between HJY or VIS alone suggests that increasing the resolution of the images might give better results than using colours in the current configuration of the II SGLC dataset, as was previously seen in Figure~\ref{fig:img_preprocessing}. The best performance was using all bands as the input, combining the higher resolution VIS images and colour information from the infrared bands. Interestingly, the results using only the Y band are similar to the ones using H, J, and Y together. Although visually more appealing, the alternative preprocessing did not always translate into better results.

Moreover, both ways of inputting the VIS band also give similar results when using it alone or when combining it with other bands. In fact, all results using a two-branch EfficientNet are compatible considering the errors, except for HJY+VIS (repeated) with the alternative preprocessing, which is only marginally better than the two band configuration with the II SLGC preprocessing when considering $1\sigma$ errors. We highlight in red the configuration used in the submission to the II SGLC, and in blue the one with the best nominal $F_{\beta}$ score.

\begin{figure*}
\centering
\begin{subfigure}{.5\textwidth}
  \centering
  \includegraphics[width=.98\linewidth]{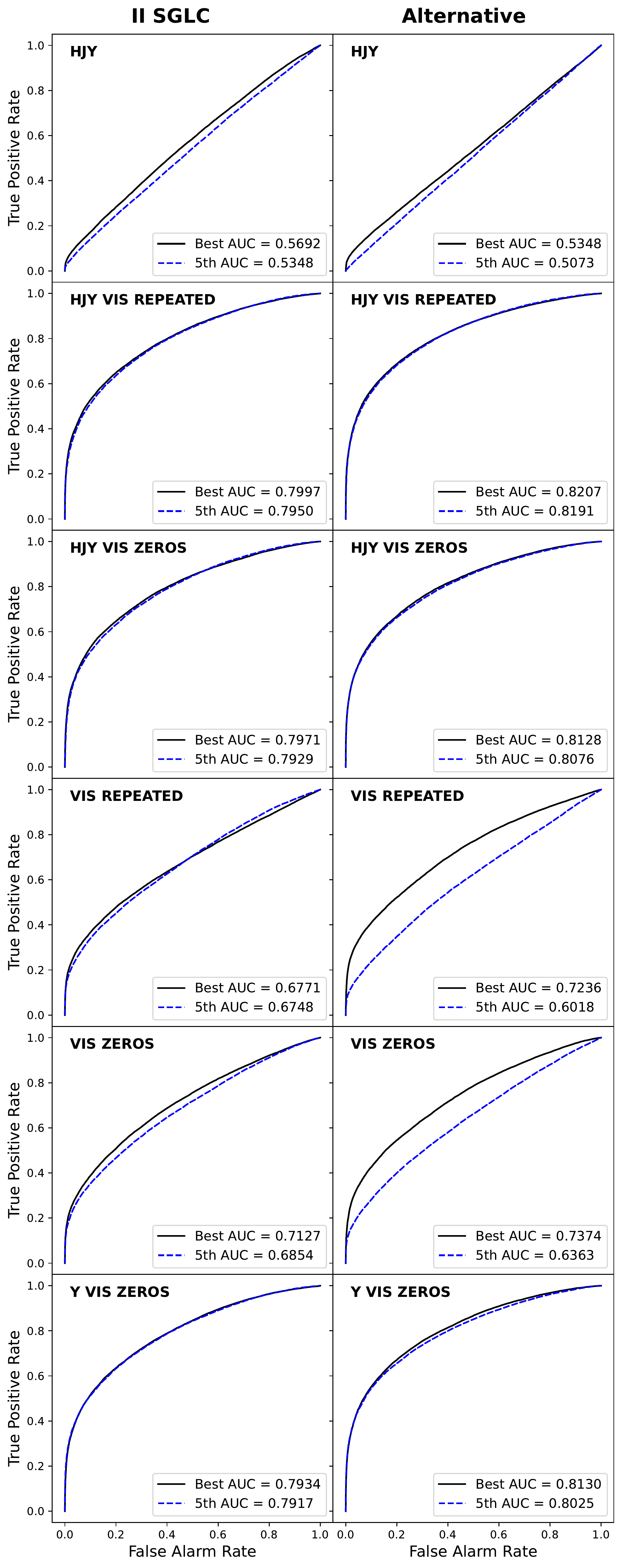}
  \caption{ROCs.}
  \label{fig:rocs_80k}
\end{subfigure}%
\begin{subfigure}{.5\textwidth}
  \centering
  \includegraphics[width=.98\linewidth]{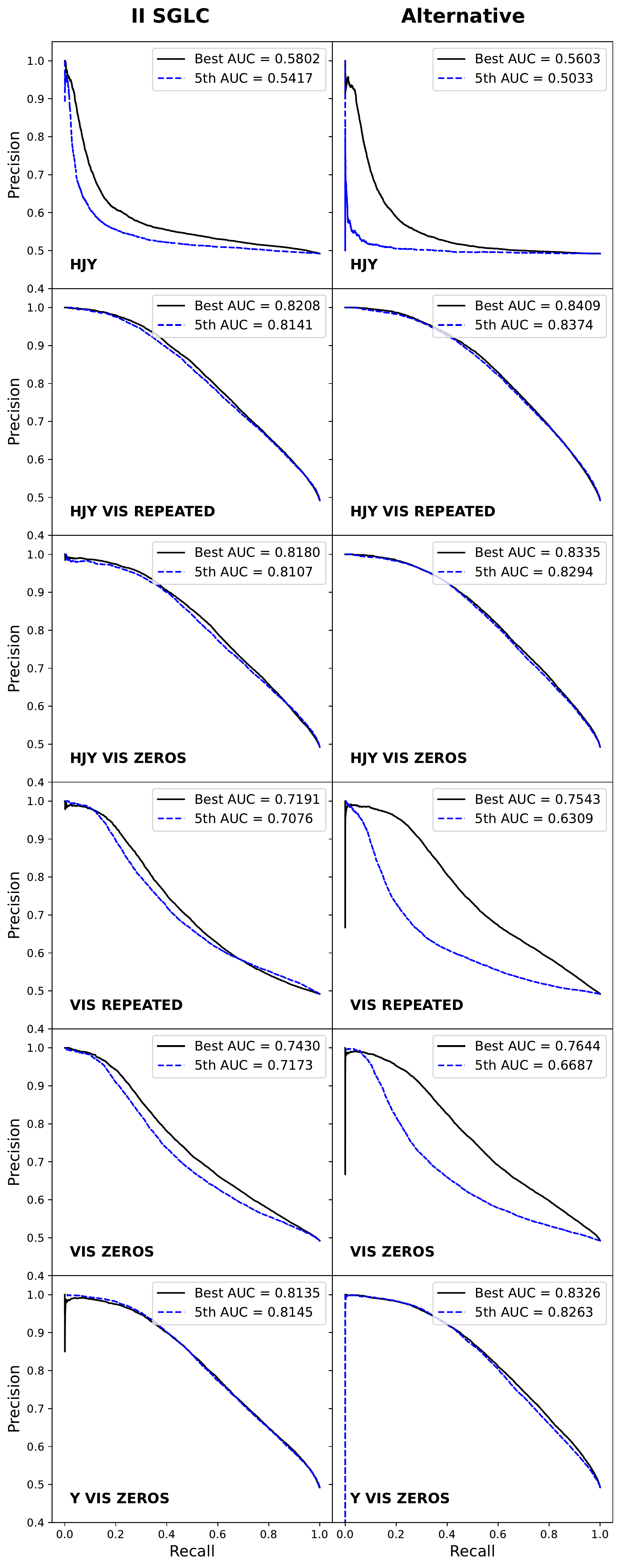}
  \caption{Precision-Recall curves.}
  \label{fig:pr_80k}
\end{subfigure}
\caption{Results of our models on the independent test group with $80,000$ objects. The best performing fold (lowest validation loss) is shown as a black solid line, and the fifth best one as a  blue dashed one.}
\label{fig:results_80k}
\end{figure*}

We also present the results for the independent test set with the resulting $80k$ images neither used in training nor validation in Figure~\ref{fig:results_80k}. We used the models with the lowest and 5th lowest validation loss from the $10$ multiple trainings for inference in this sample. The results agree with the ones reported in the table, apart from for the fold with best model using only VIS. Interestingly, we found that the best model in VIS with zeros has a better performance considering one standard deviation in the blind test set compared to the validation set results. However, the 5th best model is in agreement with table \ref{table:results}, and the AUCs for all folds confirm that the best model using VIS alone is an outlier.

\begin{table*}
\begin{tabular}{|l|lV{3.5}l|cl|cl|cl|}
\hline
 &     & \multicolumn{2}{cV{3.5}}{\textbf{\begin{tabular}[c]{@{}l@{}}No \\ Preprocessing\end{tabular}}} & \multicolumn{2}{cV{3.5}}{\textbf{\begin{tabular}[c]{@{}l@{}} II SGLC \\ Preprocessing \end{tabular}}} & \multicolumn{2}{c|}{\textbf{\begin{tabular}[c]{@{}l@{}} Alternative \\ Preprocessing\end{tabular}}} \\ \hlineB{2.5}
 
\multirow{2}{*}{\textbf{HJY}} & ROC & \multicolumn{2}{cV{3.5}}{$0.4994\pm 0.00135$} & \multicolumn{2}{cV{3.5}}{$0.5389\pm 0.01891$}  & \multicolumn{2}{c|}{$0.5350\pm 0.02414$}      \\ \cline{2-8} 
& PR  & \multicolumn{2}{cV{3.5}}{$0.4950\pm 0.0155$} & \multicolumn{2}{cV{3.5}}{$0.5535\pm 0.0210$} & \multicolumn{2}{c|}{$0.5342\pm 0.0303$} \\ \cline{2-8}
& $F_{\beta}$ & \multicolumn{2}{cV{3.5}}{$0.620002\pm 0.102507$} & \multicolumn{2}{cV{3.5}}{$0.918411\pm 0.069823$} & \multicolumn{2}{c|}{$0.800559\pm 0.098771$}    \\ \hlineB{2.5}

\multirow{2}{*}{\textbf{\begin{tabular}[c]{@{}l@{}}VIS\\ (repeated)\end{tabular}}}     
& ROC & \multicolumn{2}{cV{3.5}}{$0.5040\pm 0.0108$} & \multicolumn{2}{cV{3.5}}{$0.6589\pm 0.0423$} & \multicolumn{2}{c|}{$0.6266\pm 0.0445$}    \\ \cline{2-8} 
& PR  & \multicolumn{2}{cV{3.5}}{$0.5129\pm 0.0779$}  & \multicolumn{2}{cV{3.5}}{$0.6936\pm 0.0366$}
& \multicolumn{2}{c|}{$0.6634\pm 0.0434$} \\ \cline{2-8}
& $F_{\beta}$ & \multicolumn{2}{cV{3.5}}{$0.526797\pm 0.066389$} & \multicolumn{2}{cV{3.5}}{$0.982306\pm 0.007477$} & \multicolumn{2}{c|}{$0.979158\pm 0.011016$}    \\ \hlineB{2.5}

\multirow{2}{*}{\textbf{\begin{tabular}[c]{@{}l@{}}VIS\\ (zeros)\end{tabular}}}        
& ROC & \multicolumn{2}{cV{3.5}}{$0.5001\pm 0.0101$}  & \multicolumn{2}{cV{3.5}}{$0.7002\pm 0.0306$} & \multicolumn{2}{c|}{$0.6534\pm 0.0414$}     \\ \cline{2-8} 
& PR  & \multicolumn{2}{cV{3.5}}{$0.4884\pm 0.0384$}  & \multicolumn{2}{cV{3.5}}{$0.7326\pm 0.0256$} & \multicolumn{2}{c|}{$0.6864\pm 0.0387$}    \\ \cline{2-8}
& $F_{\beta}$ & \multicolumn{2}{cV{3.5}}{$0.548153\pm 0.066831$} & \multicolumn{2}{cV{3.5}}{$0.986763\pm 0.005173$} & \multicolumn{2}{c|}{$0.979588\pm 0.010142$}    \\ \hlineB{2.5}

\multirow{2}{*}{\textbf{\begin{tabular}[c]{@{}l@{}}HJY+VIS\\ (zeros)\end{tabular}}}    
& ROC & \multicolumn{2}{cV{3.5}}{$0.5017\pm 0.0069$}  & \multicolumn{2}{cV{3.5}}{$0.8018\pm 0.0173$} & \multicolumn{2}{c|}{$0.8147\pm 0.0161$}    \\ \cline{2-8} 
& PR  & \multicolumn{2}{cV{3.5}}{$0.4916\pm 0.0074$}  & \multicolumn{2}{cV{3.5}}{$0.8239\pm 0.0128$} & \multicolumn{2}{c|}{$0.8359\pm 0.0132$}    \\ \cline{2-8}
& $F_{\beta}$ & \multicolumn{2}{cV{3.5}}{$0.613814\pm 0.094273$} & \multicolumn{2}{cV{3.5}}{$0.990140\pm 0.003437$} & \multicolumn{2}{c|}{$0.991610\pm 0.003873$}    \\ \hlineB{2.5}

\multirow{2}{*}{\textbf{\begin{tabular}[c]{@{}l@{}}HJY+VIS\\ (repeated)\end{tabular}}} 
& ROC & \multicolumn{2}{cV{3.5}}{$0.4932\pm 0.0102$}  & \multicolumn{2}{cV{3.5}}{$0.8016\pm 0.0186$} & \multicolumn{2}{c|}{{\color{blue}$0.8295\pm 0.0088$}}    \\ \cline{2-8} 
& PR  & \multicolumn{2}{cV{3.5}}{$0.4921\pm 0.0175$}  & \multicolumn{2}{cV{3.5}}{$0.8230\pm 0.0149$} & \multicolumn{2}{c|}{{\color{blue}$0.8469\pm 0.0081$}}    \\ \cline{2-8}
& $F_{\beta}$ & \multicolumn{2}{cV{3.5}}{$0.619433\pm 0.106441$} & \multicolumn{2}{cV{3.5}}{$0.990402\pm 0.004087$} & \multicolumn{2}{c|}{{\color{blue}$0.992070\pm 0.003427$}}    \\ \hlineB{2.5}

\multirow{2}{*}{\textbf{\begin{tabular}[c]{@{}l@{}}Y+VIS\\ (zeros)\end{tabular}}}      
& ROC & \multicolumn{2}{cV{3.5}}{$0.5011\pm 0.0174$}  & \multicolumn{2}{cV{3.5}}{\textcolor{red}{$0.8015\pm 0.0186$}} & \multicolumn{2}{c|}{$0.8149\pm 0.0144$}    \\ \cline{2-8} 
& PR  & \multicolumn{2}{cV{3.5}}{$0.4939\pm 0.0124$}  & \multicolumn{2}{cV{3.5}}{\textcolor{red}{$0.8243\pm 0.0135$}} & \multicolumn{2}{c|}{$0.8369\pm 0.0120$}    \\ \cline{2-8}
& $F_{\beta}$ & \multicolumn{2}{cV{3.5}}{$0.613662\pm 0.125263$} & \multicolumn{2}{cV{3.5}}{\textcolor{red}{$0.989978\pm 0.004941$}} & \multicolumn{2}{c|}{$0.990235\pm 0.004237$}    \\ \hlineB{2.5}

\end{tabular}
\caption{Area under curve (AUC) for ROC and Precision-Recall (PR) curves, and the $F_{\beta}$ score for different bands and different preprocessing, on a reduced sample (20,000 images), using a cross-validation procedure. The scores are calculated on a blind test sample comprising $10\%$ of this reduced sample. As per the challenge, we use $\beta^2 = 0.001$. The combination submitted to the challenge is highlighted in red, and the best one in blue.}
\label{table:results}
\end{table*}

\subsection{Strong Lensing classification interpretation}
\label{sec:lime}
Deep Learning predictions are often hard to interpret and obtain intuition. Moreover, there is no straightforward standard procedure to justify the algorithm choices. One of the reasons for that is the complexity of those models, including the number of free parameters and sparsity. Therefore, several techniques were proposed to infer and interpret the outputs from a given Network; among them, we chose a popular approach named Local Interpretable Model-Agnostic Explanations \cite[LIME;][]{lime}. This is a framework that can be applied to several machine learning methods as has been applied in several ML pipelines \citep[see, e.g.][]{SLime, lime_music, cancer_lime}. The method makes perturbations in small parts of the input and evaluates how the predictions change. It treats the model as a complete black-box (hence the name \textit{model agnostic}), approximating it locally by a linear model, which is simpler to analyze compared to the global model.
\par For image classification, LIME does this by running the model several times, perturbing different image regions, and checking the predictions. The superpixels (group of adjacent pixels) defined by the LIME technique are then classified according to their importance to the classification.
\par
\begin{figure*}
    \centering
    \includegraphics[width=\textwidth]{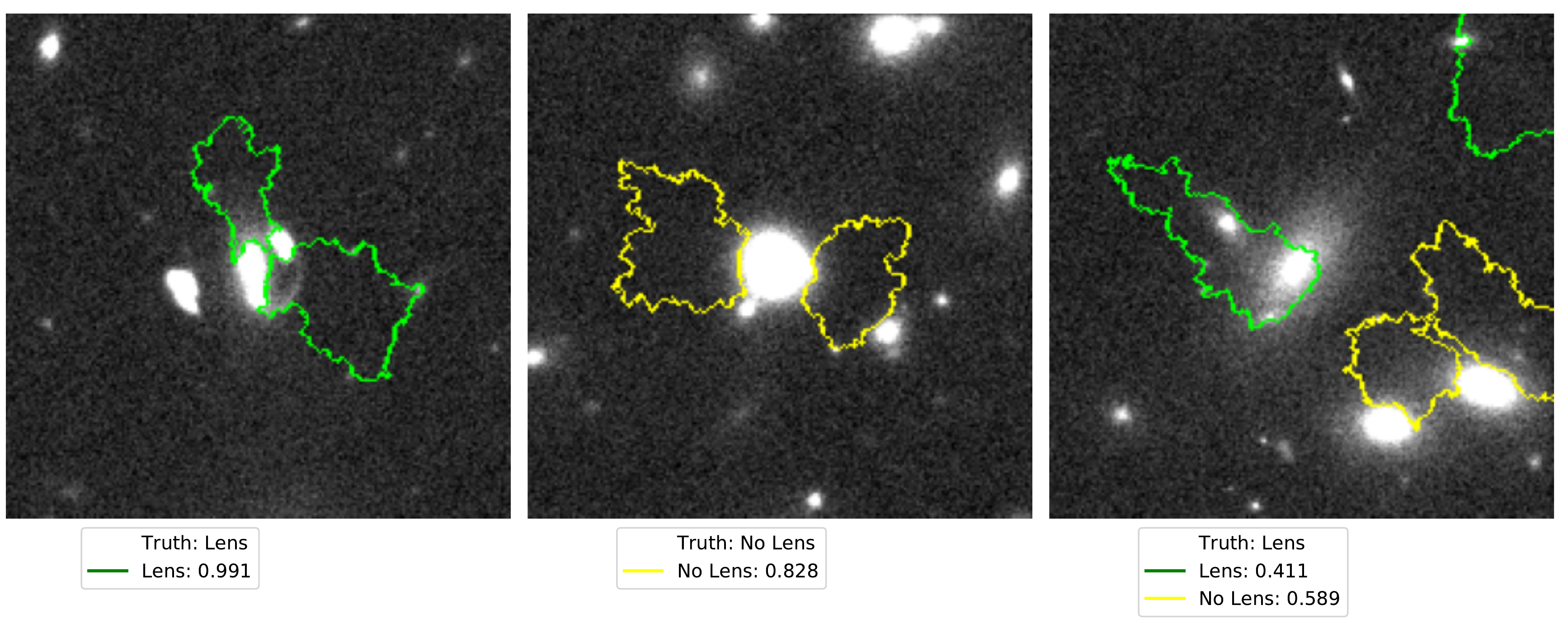}
    \caption{LIME analysis for three images in the dataset. Highlighted regions correspond to the two most important superpixels when determining the classification, green for lenses and yellow for non-lenses. The predictions of the network and the ground truth are also shown.}
    \label{fig:lime}
\end{figure*}

 We applied LIME to our sample images and made a visual assessment of what are the superpixels relevant for its classification. In Figure \ref{fig:lime} we present three example images from our dataset with their respective predictions and ground truth, with the two most important superpixels highlighted. In the left and middle panels, the probabilities are fairly high for one of the classes, so we show only the superpixels important to that class, while on the right panel, we show the regions important for both classes since the probabilities are close. It is worth noticing that the model searched around the central galaxy for the lens and gave high probabilities for being a lens or not based on finding it, which agrees with the intuition on how a human classifier would analyze the image. However, when the field is crowded with bright objects, it searches around background galaxies for lenses, finding none. This suggests that Deep Learning algorithms might be misguided in crowded fields. Nevertheless, since it also searched around the central galaxy for the lens, both probabilities are comparable.

\section{Detection limits}
\label{sec:limits}

\begin{figure}
    \centering
    \includegraphics[width=\linewidth]{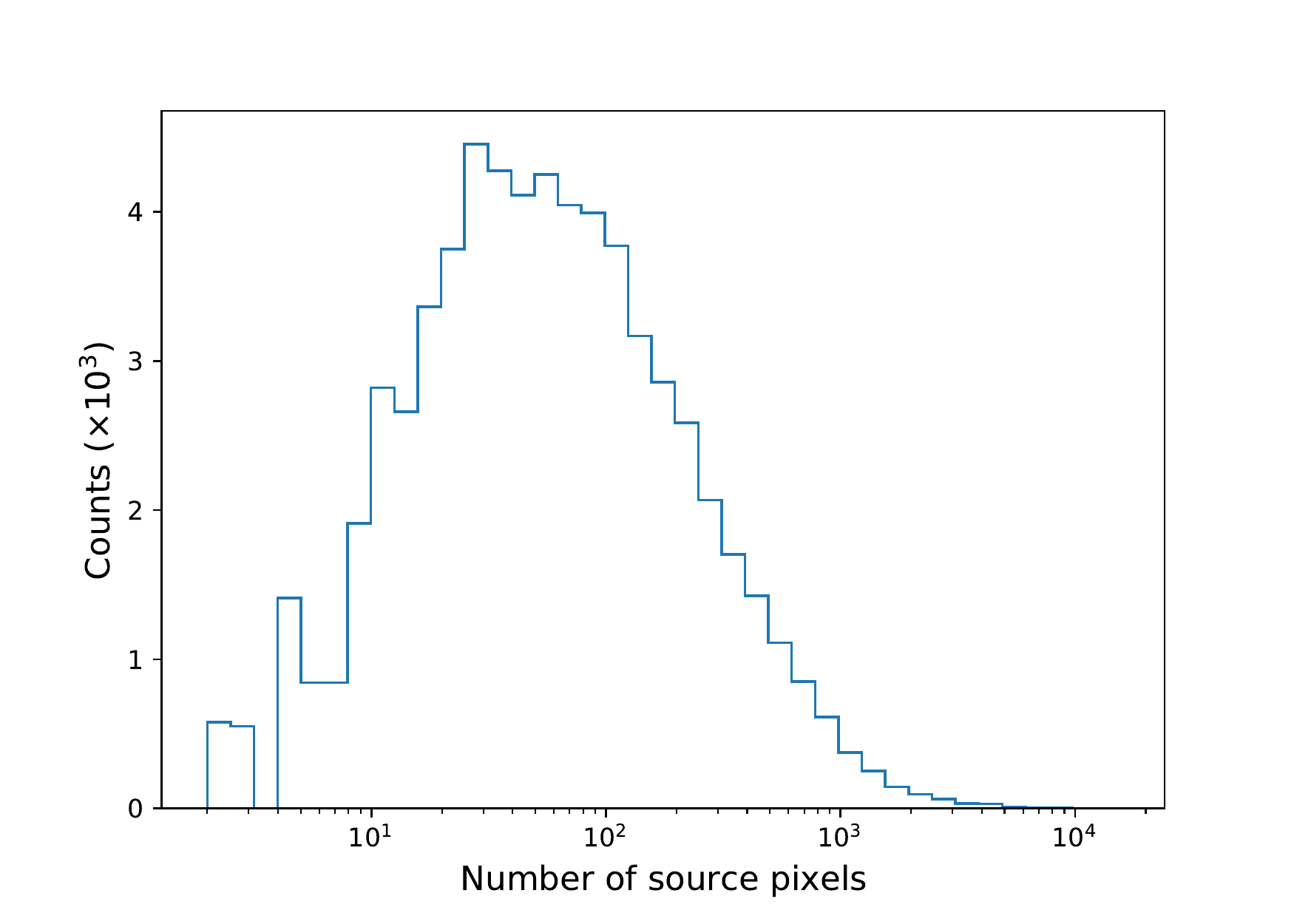}
    \caption{Distribution of number of source pixels above the background for images satisfying both other criteria for being considered a lens (effective magnification and number of groups of sources).}
    \label{fig:hist_npix}
\end{figure}

Conceptually,  a Strong Lensing system is defined when the light of a given source is strongly deflected by the lens. However, the strict definition of whether the system is detectable or not is not clearly defined, depending on the survey definitions, observational conditions, and instrument configurations. For instance, the simulated images can have just a few source pixels above the background level so that even if there is a signal, it might be undetectable. The II SGLC organizers gave their own definition of detectable Strong Lensing, which we reproduce in order to obtain a truth table of SL and not SL, as presented in Section 2.1. The experience from I SGLC showed that human visual inspection is less sensitive than automated DL algorithms in simulations. Thus, we investigate the detection limits of our method, in particular, how the number of source pixels above $1\sigma$ of the background level, $N_{pix}$, which was used as a criterion to define SL in the II SGLC, can affect our DL model's performance. Figure \ref{fig:hist_npix} shows the distribution of $N_{pix}$ for images that fulfill the other two criteria for lenses, all with $N_{pix}>0$ (since, in principle, any image with at least one source pixel above the background level could be considered a SL system). Even though the challenge set a minimum of $20$, the median of the distribution is approximately $50$, with still a relevant number of objects with $N_{pix}>100$.

\begin{figure*}
    \centering
    \includegraphics[width=\linewidth]{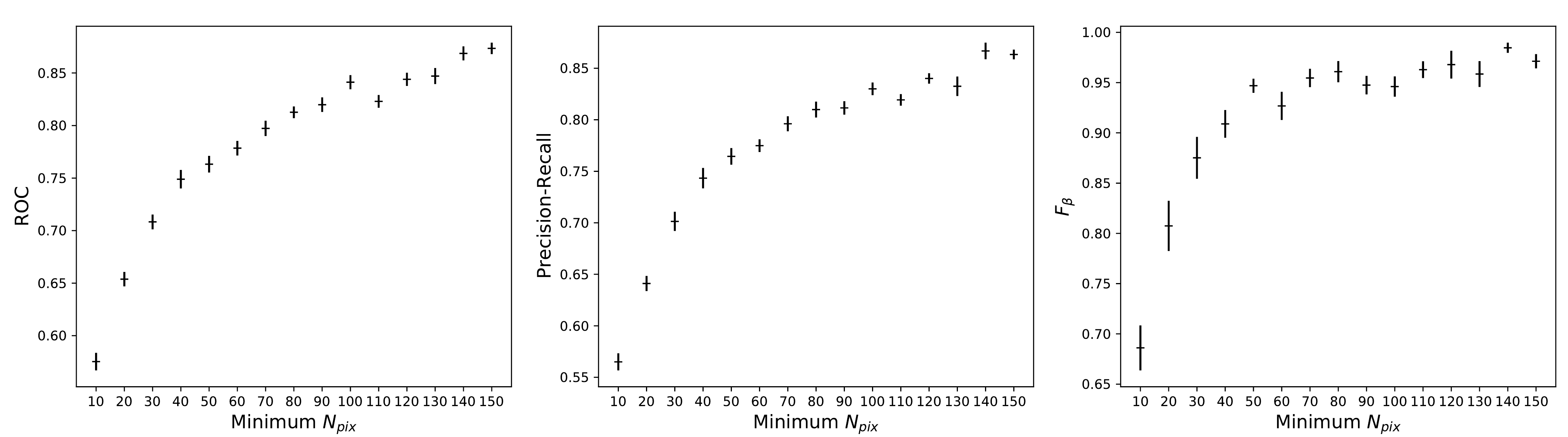}
    \caption{Metrics for different minimum values of $N_{pix}$. The points correspond to the minimum, and the error bars to one standard deviation considering all models.}
    \label{fig:metrics_npix}
\end{figure*}

In order to assess the performance of our model in more adverse conditions, we test our trained HJY+VIS (repeated) models in test sets composed of lensed systems in a given $N_{pix}$ range, fulfilling the other criteria mentioned in \S\ref{sec:dataexp}. This set is complemented with the non-lenses category defined, for the purposes of this test only, by an equal number of objects with $N_{pix}$ less than the minimum value for that given range. All these objects are taken from the $80,000$ objects that we used as an independent test set (see \S\ref{sec:performance}). We start with. $N_{pix}=10$ and go to $150$ in steps of 10; for example, in the first range the lensed systems have $10\leq N_{pix}<20$ and non-lensed have $N_{pix}<10$. We use the results of our ten trained models (one for each fold in the cross-validation scheme) for each range to obtain the error bars. Figure \ref{fig:metrics_npix} show the mean and one standard deviation ROC AUC, Precision-Recall AUC, and $F_{\beta}$ considering the results from all ten models. For $N_{pix}<30$, the net is bot much better from random guessing, while for $N_{pix}\geq 70$ we start seeing results comparable to the inference on the full sample.

\section{Adaptability}
\label{sec:adaptability}
Most of the DL algorithms are tailor-made for a specific dataset. This scenario is reasonable in a context of a data competition. However, the pursuance of an adaptable algorithm can save lots of development in new datasets making the efforts to analyze new data more efficient and accessible, not requiring a great amount of time from DL experts, increasing the model usage capability and relevance to the community. DL models are usually data-hungry, relying upon massive simulated datasets in specific surveys, which are not always available or convenient, as simulations will focus on a range of parameters and models. Additionally, adapting the method for a different scenario also validates the methodology. 

The dataset suitable to evaluate whether the DL is adaptable should be considered uniform in a given survey condition that is different from the dataset used in the initial training. It also needs to be abundant enough to determine the algorithm's performance and how many training samples are required to make a fine-tuning. Therefore, we evaluate the use of a different dataset, namely, the one used in I SGLC in the current trained algorithm. Differently from the II SGLC data, based in Euclid, space-based conditions, the main multiband I SGLC dataset was built to represent ground-based images using the ESO's VLT Survey Telescope in KIDS-like data. This included the observational conditions of the KIDS survey, the level of noise, and the bands. The dataset, same used as training in I SGLC,  presents a total of $20,000$ systems with a truth table among lenses and not lenses in four bands $u,g,r,i$. Due to the nature of the I SGLC simulations, the dataset is relevant to highlight performance in different surveys/observational conditions. However, they are simulated with the same kind of SL lensing algorithms, making it not suitable to determine the limitations of specific SL modeling methods.

We start our investigation by applying the trained in II SGLC using the HJY+VIS network in the new KIDS-like data. After some initial tests, we found that the best way to split the four bands of this data to fit our $3+1$ scheme was to leave the $r$ band alone, combining $u$, $g$ and $i$. The images were preprocessed and adjusted using the same kind of procedure employed in the II SGLC. However, the images had a different size in pixels to the Euclid-like dataset used to train the models, so we resize them using a standard routine from the OpenCV library \citep{opencv_library}. The performance of the model when trying to make inference directly in this new dataset was consistent with a random guess, i.e., AUC of ROC $\sim 0.5$. In order to improve this result, we select a small number of samples in the new data and use them for training the model, leaving every weight in the net free. The cross-validation procedure was the following: first, we split the initial $20,000$ systems into $10$ groups of $2,000$. Each group will be used as an independent test. For each test group, we select a given number of images ($N_{img}$) from the $18,000$ remaining ones for training and the same number for validation. 

\begin{figure*}
    \centering
    \includegraphics[width=\linewidth]{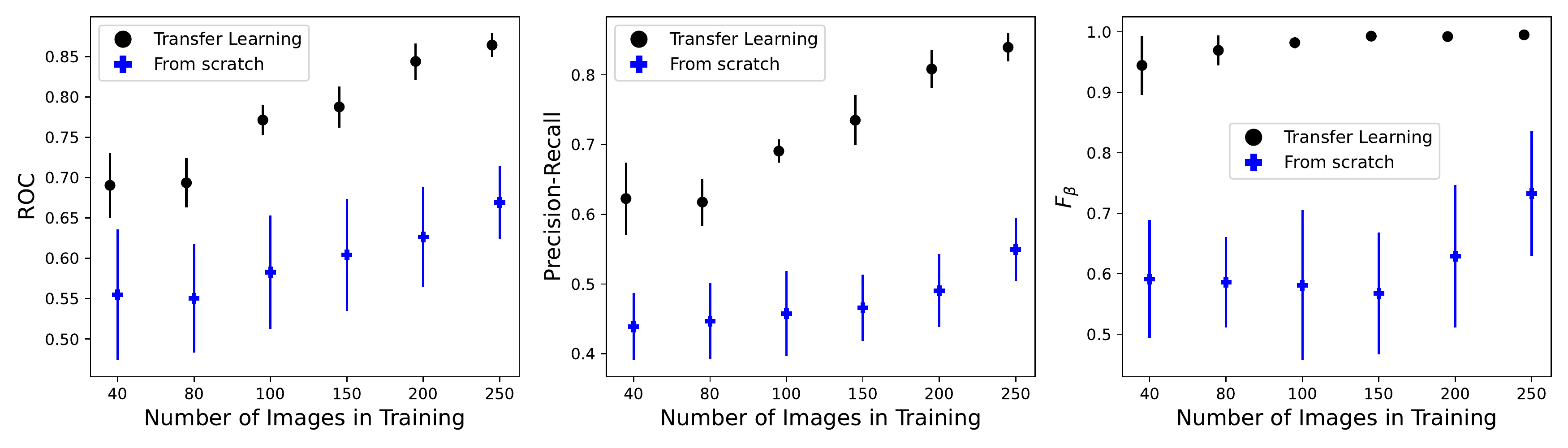}
    \caption{Metrics for our model tested on I SLGC data, retraining one of our models (Transfer learning, black circles) and using random weights (from scratch, blue crosses). We use an increasing number of images for training, and report one standard deviation errors for the cross validation. See text for details.}
    \label{fig:metrics_challenge1}
\end{figure*}

In figure \ref{fig:metrics_challenge1} we present our metrics for this test using different numbers of training images. In black, we report the results using the trained model in the II SGLC; for comparison, we also show in blue the same metrics in the same number of images, now using a network trained from scratch, i.e., initialized with random weights. We see that the trained model already has an AUC ROC above the randomness level with $40$ images even though the model trained from scratch is still guessing. With  $200$ images, we see that the results are comparable to the ones obtained with the II SGLC dataset. It can also be seen that in all cases, retraining the model gives better and more consistent results.

\section{Discussion and Concluding remarks}
\label{sec:conclusions}
\subsection{Summary}
In this contribution, we present a pipeline and the strategy employed to obtain the highest performance result in terms of $f_{\beta}$ score of the II Strong Gravitational Lensing Challenge. We discuss the choices in the code, architecture definition, how to work with images with different resolutions, the importance of a preprocessing evaluating it by the code performance and visual assessment. We use the Local Interpretable Model-Agnostic Explanations technique to infer the relevant features used in the classification. We also present a prescription on how to adapt the pipeline for a different dataset and the detectability limit in terms of lensed pixels. Finally, we made a public release of the code, including neural network weights. In the next paragraphs, we summarize the lessons learned.

\subsection{Data Visualization, preprocessing and normalization}

The sample's visual inspection shows that just normalizing the images without a proper contrast adjustment, for instance, clipping the histogram, makes us see no lensing feature at all. This was later confirmed also by the Deep Learning pipeline. However, by comparing the two preprocessing we use, even if we perceive a visual gain, this does not necessarily reflect into better performance, considering the $1\sigma$ errors. Therefore, we conclude that preprocessing the images in order to make them visually convincing as lenses is an important step. However, fine tunning might be more critical to visual assessments than to DL algorithms.

\subsection{Multiple bands and resolution relevance}

There are multiple sources of spurious detections in SL analysis, and some of them are just image artifacts. Nonetheless, others are images that might look like a lens, as in the case of edge-on galaxies and some spiral galaxies. One popular approach to reduce this issue is to use colours and look for red galaxies with a blue object close. The colour information is considered an important feature to find lenses \citep[see, for instance, ][]{10.1093/mnrasl/slx173,10.1093/mnras/sty1923}. The use of colour queries is considered at least an interesting way to make a preselection and improve the final Deep Learning result purity. In fact, even in the I SGLC results the single band scenario found lower performance in terms of the considered metric, the ROC AUC.
The II SGLC in a Euclid-like scenario allowed to make Deep Learning classifications using images with colour information in multiple bands but also higher resolution images. Interestingly the networks using only HJY bands did not find a competitive fit. On the other hand, using the VIS band only with a higher resolution found better results than HJY, which is close to a random guess. Still, the VIS band only has inferior results if compared to the runs using all information. This result suggests that the use of high-resolution images might play an important role and was in fact, more relevant than the multiple bands in the cases tested in this contribution. It is worth mentioning that the threshold defined by the maximum $F_{\beta}$, with $\beta=0.001$, privileged a pure sample instead of a complete one, obtaining in our best network, HJY+VIS with alternative preprocessing, around  $\sim 99\%$ purity with completeness of around $45\%$. This choice is justified for the same reason a preselection of targets is implemented by colour queries and other methods to reduce the number of nonlenses in a survey where we have billions of not SL for thousands of SL.

\subsection{Underfitting and Overfitting}

During the network definition and training process, we found, on several occasions, underfitting results, i.e., training losses were not minimized, and also overfitting, where the discrepancy between training and validation losses increases with the epochs. The underfitting was found mainly when using no preprocessing, just a simple data normalization; this result is presented in Fig. \ref{fig:diff_loss}. It also agrees with the fact, already mentioned, that without the preprocessing no features were visible to human eyes. 

Nonetheless, we also find overfitting in many of our tests. Even the best configurations presented a tendency to overfit if we just let the training proceed unbounded. Rather than choose an early stopping method that might miss a big picture of training by quickly interrupting it to avoid overfitting, we deal with that by just choosing the epoch where the validation loss was the lowest to perform our evaluation of the model. This was an important strategy to the competition, enabling us to choose the network weights that could better generalize to the validation sample. Interestingly, when we delivered our submission to the II SGLC, the higher performance network was not the one that best performed in terms of $F_{\beta}$ in the testing sample, but precisely the ones that generalized better in the validation sample. This suggests that, in order to adapt for this blind sample, the best strategy was to choose the one with the lowest validation loss. A more detailed discussion on the network entries is a topic to be presented in the II SGLC results paper. 

Additionally to the observed tendency to overfitting, we also found unexpected overfitting in the very early epochs, i.e., from epoch $2$ by using the Tensorflow native implementation of EfficientNets. This was an overfitting so early that happened before the validation loss dropped a level where we could find a classification better than random guess. After some investigation, we find that there are some differences between the TensorFlow implementation and ours, based on the original EfficientNet GitHub \footnote{https://github.com/qubvel/efficientnet}, where the former incorporates into the architecture some preprocessing layers. Since our data was already fully prepared, these extra steps caused the issues mentioned.

\subsection{Neural Network Complexity}

From the computer vision challenges \citep{ILSVRC15} experience, the use of deeper networks with similar architectures in terms of layers and structure are usually high-performing if we are able to train successfully, with no strong overfitting. This scenario changes when some innovation on the architecture is presented, like the Resnet \citep{he2016deep}. This effect was also presented in the EfficientNet paper. The scaled networks, with a bigger number of parameters, presented a better performance compared to the ones with lower parameters, this effect was more evident with the initial B0-B4 than the laters up to B7 where the performance gain was smaller \citep[for further details, see Figure 1 of][]{efficientnet}. Nevertheless, these deeper networks require more computation time, high-performing hardware and the gain in performance is not always sufficient to justify this choice depending on the specific problem, particularly in a cross-validation scheme where we evaluate the variability of the results for a given train/validation sets. In our experiments with SL we found this scenario with the EfficientNets with a great number of free parameters, and we choose the B2 architecture as a trade-off between computation time efficiency and performance.

\subsection{Deep Learning Strong Lensing finding decision making}
There is not a general interpretation theory for DL decision-making. In fact this is an active research topic \citep[see, e.g., ][]{DL_interpretability}. Thus, we made use of LIME technique to infer what is more important to the DL decision making process. This is not the only possible choice, for instance, in a recent paper by \citet{2022arXiv220212776W}, the authors used a different set of techniques to perform assessment of a Strong Lensing finder and found that it in high confidence lens their CNN model highlights the arc or ring shapes. Our results reveal that the important regions for classification are the edges of the central galaxy, which agrees with the visual inspection intuition. We noticed that crowded regions were influenced by the neighboring objects. This might be an important issue for lenses with multiple images but no pronounced arcs and regions close to the center of clusters. One possibility is to work with smaller images. In principle, it is worth noticing that one could work with the same trained network but using some stamps with smaller sizes by resizing the images and with a fine-tuning. This approach was implemented successfully when we adapted the network for the I SGLC sample, where the original images had $101 \times 101$ pixels and the $r$ band was resized to fit the VIS network branch where we had $200\times200$ images.

\subsection{Generic Strong Lensing finders based on Deep Learning models are possible?}

The deep learning algorithms are usually tailor-made for a specific survey. Commonly this means enormous efforts to define a suitable high-performing architecture and relies in massive simulations to train the models, which are ideally made to represent the expected population of SL. Here, we presented a simple prescription to adapt the pipeline to different survey conditions, where one can use a small amount of data. This could be used to fine-tuning by training in real data, which is usually very limited and avoid the need to produce new Strong Lensing finders from scratch. Depending on the specific goal of the SL search, this approach has to be used carefully, as the known sample of real SL is highly inhomogeneous and subject to complex selection functions \citep[see, for instance, ][]{2022zndo...5836022G}. However, this might be of particular interest to specialize the algorithm to a specific population of rare systems. The two branched architecture presented here to deal with images of different resolutions also offers a possible path to integrate multiple data from current and future multiband surveys.


\section*{Data Availability}
We make the Deep Learning algorithms publicly available in \url{https://github.com/cdebom/cast_lensfinder}. The trained Deep Learning models are also public and can be downloaded in \url{https://doi.org/10.5281/zenodo.6344064}. The image datasets developed in the context of the I and II Strong Gravitational Lensing Challenge are available at \url{http://metcalf1.difa.unibo.it/blf-portal/gg_challenge.html}

\section*{Acknowledgements}


The authors would like to thank to the Bologna Lens Factory team  for organizing the II Strong Gravitational Lensing  Challenge and producing the full datasets used in this work. CRB acknowledges the financial support from CNPq (316072/2021-4). CF acknowledges the financial support from CNPq (processes 433615/2018-4 and 314672/2020-6).
The authors acknowledge the LITCOMP/COTEC/CBPF multi-GPU development team for all the support in the Artificial Intelligence infrastructure and Sci-Mind's High-Performance multi-GPU system.



\clearpage

\newpage
\clearpage


\newpage
\clearpage


\bibliographystyle{mnras} 
\bibliography{bibliografia}

\begin{thebibliography}{}
\makeatletter
\relax
\def\mn@urlcharsother{\let\do\@makeother \do\$\do\&\do\#\do\^\do\_\do\%\do\~}
\def\mn@doi{\begingroup\mn@urlcharsother \@ifnextchar [ {\mn@doi@}
  {\mn@doi@[]}}
\def\mn@doi@[#1]#2{\def\@tempa{#1}\ifx\@tempa\@empty \href
  {http://dx.doi.org/#2} {doi:#2}\else \href {http://dx.doi.org/#2} {#1}\fi
  \endgroup}
\def\mn@eprint#1#2{\mn@eprint@#1:#2::\@nil}
\def\mn@eprint@arXiv#1{\href {http://arxiv.org/abs/#1} {{\tt arXiv:#1}}}
\def\mn@eprint@dblp#1{\href {http://dblp.uni-trier.de/rec/bibtex/#1.xml}
  {dblp:#1}}
\def\mn@eprint@#1:#2:#3:#4\@nil{\def\@tempa {#1}\def\@tempb {#2}\def\@tempc
  {#3}\ifx \@tempc \@empty \let \@tempc \@tempb \let \@tempb \@tempa \fi \ifx
  \@tempb \@empty \def\@tempb {arXiv}\fi \@ifundefined
  {mn@eprint@\@tempb}{\@tempb:\@tempc}{\expandafter \expandafter \csname
  mn@eprint@\@tempb\endcsname \expandafter{\@tempc}}}

\bibitem[\protect\citeauthoryear{Abadi et~al.,}{Abadi
  et~al.}{2016}]{tensorflow}
Abadi M.,  et~al., 2016, CoRR, abs/1603.04467

\bibitem[\protect\citeauthoryear{{Abdelsalam}, {Saha}  \&
  {Williams}}{{Abdelsalam} et~al.}{1998}]{1998MNRAS.294..734A}
{Abdelsalam} H.~M.,  {Saha} P.,   {Williams} L.~L.~R.,  1998, \mn@doi [\mnras]
  {10.1046/j.1365-8711.1998.01356.x}, \href
  {http://adsabs.harvard.edu/abs/1998MNRAS.294..734A} {294, 734}

\bibitem[\protect\citeauthoryear{{Akhshik} et~al.,}{{Akhshik}
  et~al.}{2020}]{2020ApJ...900..184A}
{Akhshik} M.,  et~al., 2020, \mn@doi [\apj] {10.3847/1538-4357/abac62}, \href
  {https://ui.adsabs.harvard.edu/abs/2020ApJ...900..184A} {900, 184}

\bibitem[\protect\citeauthoryear{Avestruz, Li, Zhu, Lightman, Collett  \&
  Luo}{Avestruz et~al.}{2019}]{avestruz2019automated}
Avestruz C.,  Li N.,  Zhu H.,  Lightman M.,  Collett T.~E.,   Luo W.,  2019,
  The Astrophysical Journal, 877, 58

\bibitem[\protect\citeauthoryear{{Bartelmann}, {Huss}, {Colberg}, {Jenkins}  \&
  {Pearce}}{{Bartelmann} et~al.}{1998}]{1998A&A...330....1B}
{Bartelmann} M.,  {Huss} A.,  {Colberg} J.~M.,  {Jenkins} A.,   {Pearce} F.~R.,
   1998, \aap, \href {http://adsabs.harvard.edu/abs/1998A\%26A...330....1B}
  {330, 1}

\bibitem[\protect\citeauthoryear{Bayer, Chatterjee, Koopmans, Vegetti, McKean,
  Treu  \& Fassnacht}{Bayer et~al.}{2018}]{bayer2018observational}
Bayer D.,  Chatterjee S.,  Koopmans L.,  Vegetti S.,  McKean J.,  Treu T.,
  Fassnacht C.,  2018, arXiv preprint arXiv:1803.05952

\bibitem[\protect\citeauthoryear{{Bayliss}}{{Bayliss}}{2012}]{2012ApJ...744..156B}
{Bayliss} M.~B.,  2012, \mn@doi [\apj] {10.1088/0004-637X/744/2/156}, \href
  {http://adsabs.harvard.edu/abs/2012ApJ...744..156B} {744, 156}

\bibitem[\protect\citeauthoryear{{Belokurov}, {Evans}, {Hewett}, {Moiseev},
  {McMahon}, {Sanchez}  \& {King}}{{Belokurov}
  et~al.}{2009}]{2009MNRAS.392..104B}
{Belokurov} V.,  {Evans} N.~W.,  {Hewett} P.~C.,  {Moiseev} A.,  {McMahon}
  R.~G.,  {Sanchez} S.~F.,   {King} L.~J.,  2009, \mn@doi [\mnras]
  {10.1111/j.1365-2966.2008.14075.x}, \href
  {http://adsabs.harvard.edu/abs/2009MNRAS.392..104B} {392, 104}

\bibitem[\protect\citeauthoryear{{Bom}, {Furlanetto}, {More}, {Brandt},
  {Makler}  \& {Santiago}}{{Bom} et~al.}{2015}]{2015mgm..conf.2088D}
{Bom} C.~R.,  {Furlanetto} C.,  {More} A.,  {Brandt} C.,  {Makler} M.,
  {Santiago} B.,  2015, in {Rosquist} K.,  ed., Thirteenth Marcel Grossmann
  Meeting: On Recent Developments in Theoretical and Experimental General
  Relativity, Astrophysics and Relativistic Field Theories. pp 2088--2090,
  \mn@doi{10.1142/9789814623995_0364}

\bibitem[\protect\citeauthoryear{{Bom}, {Makler}, {Albuquerque}  \&
  {Brandt}}{{Bom} et~al.}{2017}]{2016arXiv160704644B}
{Bom} C.~R.,  {Makler} M.,  {Albuquerque} M.~P.,   {Brandt} C.~H.,  2017,
  \mn@doi [\aap] {10.1051/0004-6361/201629159}, \href
  {http://adsabs.harvard.edu/abs/2017A\%26A...597A.135B} {597, A135}

\bibitem[\protect\citeauthoryear{{Bom}, {Poh}, {Nord}, {Blanco-Valentin}  \&
  {Dias}}{{Bom} et~al.}{2019}]{Bom2019}
{Bom} C.,  {Poh} J.,  {Nord} B.,  {Blanco-Valentin} M.,   {Dias} L.,  2019,
  arXiv e-prints, \href {https://ui.adsabs.harvard.edu/abs/2019arXiv191106341B}
  {p. arXiv:1911.06341}

\bibitem[\protect\citeauthoryear{{Bom} et~al.,}{{Bom}
  et~al.}{2021}]{2021MNRAS.507.1937B}
{Bom} C.~R.,  et~al., 2021, \mn@doi [\mnras] {10.1093/mnras/stab1981}, \href
  {https://ui.adsabs.harvard.edu/abs/2021MNRAS.507.1937B} {507, 1937}

\bibitem[\protect\citeauthoryear{Bradski}{Bradski}{2000}]{opencv_library}
Bradski G.,  2000, Dr. Dobb's Journal of Software Tools

\bibitem[\protect\citeauthoryear{{Cabanac} et~al.,}{{Cabanac}
  et~al.}{2007}]{2007A&A...461..813C}
{Cabanac} R.~A.,  et~al., 2007, \mn@doi [\aap] {10.1051/0004-6361:20065810},
  \href {http://adsabs.harvard.edu/abs/2007A\%26A...461..813C} {461, 813}

\bibitem[\protect\citeauthoryear{{Carrasco} et~al.,}{{Carrasco}
  et~al.}{2010}]{2010ApJ...715L.160C}
{Carrasco} E.~R.,  et~al., 2010, \mn@doi [\apjl]
  {10.1088/2041-8205/715/2/L160}, \href
  {http://adsabs.harvard.edu/abs/2010ApJ...715L.160C} {715, L160}

\bibitem[\protect\citeauthoryear{Cheng Keyang Wang~Ning}{Cheng Keyang
  Wang~Ning}{2020}]{DL_interpretability}
Cheng Keyang Wang~Ning Shi~Wenxi Z.~Y.,  2020, \mn@doi [Journal of Computer
  Research and Development] {10.7544/issn1000-1239.2020.20190485}, 57, 1208

\bibitem[\protect\citeauthoryear{Cheng, Li, Conselice, Arag{\'o}n-Salamanca,
  Dye  \& Metcalf}{Cheng et~al.}{2020}]{cheng2020identifying}
Cheng T.-Y.,  Li N.,  Conselice C.~J.,  Arag{\'o}n-Salamanca A.,  Dye S.,
  Metcalf R.~B.,  2020, Monthly Notices of the Royal Astronomical Society, 494,
  3750

\bibitem[\protect\citeauthoryear{{Coe}, {Ben{\'{\i}}tez}, {Broadhurst}  \&
  {Moustakas}}{{Coe} et~al.}{2010}]{2010ApJ...723.1678C}
{Coe} D.,  {Ben{\'{\i}}tez} N.,  {Broadhurst} T.,   {Moustakas} L.~A.,  2010,
  \mn@doi [\apj] {10.1088/0004-637X/723/2/1678}, \href
  {http://adsabs.harvard.edu/abs/2010ApJ...723.1678C} {723, 1678}

\bibitem[\protect\citeauthoryear{{Collett}}{{Collett}}{2015}]{2015ApJ...811...20C}
{Collett} T.~E.,  2015, \mn@doi [\apj] {10.1088/0004-637X/811/1/20}, \href
  {http://adsabs.harvard.edu/abs/2015ApJ...811...20C} {811, 20}

\bibitem[\protect\citeauthoryear{{Cooray}}{{Cooray}}{1999}]{1999A&A...341..653C}
{Cooray} A.~R.,  1999, \aap, \href
  {http://adsabs.harvard.edu/abs/1999A26A...341..653C} {341, 653}

\bibitem[\protect\citeauthoryear{{Despali}, {Vegetti}, {White}, {Giocoli}  \&
  {van den Bosch}}{{Despali} et~al.}{2018}]{2018MNRAS.475.5424D}
{Despali} G.,  {Vegetti} S.,  {White} S. D.~M.,  {Giocoli} C.,   {van den
  Bosch} F.~C.,  2018, \mn@doi [\mnras] {10.1093/mnras/sty159}, \href
  {https://ui.adsabs.harvard.edu/\#abs/2018MNRAS.475.5424D} {475, 5424}

\bibitem[\protect\citeauthoryear{Diehl et~al.,}{Diehl
  et~al.}{2017}]{Diehl_2017}
Diehl H.~T.,  et~al., 2017, \mn@doi [The Astrophysical Journal Supplement
  Series] {10.3847/1538-4365/aa8667}, 232, 15

\bibitem[\protect\citeauthoryear{{Ebeling}, {Stockmann}, {Richard}, {Zabl},
  {Brammer}, {Toft}  \& {Man}}{{Ebeling} et~al.}{2018}]{2018ApJ...852L...7E}
{Ebeling} H.,  {Stockmann} M.,  {Richard} J.,  {Zabl} J.,  {Brammer} G.,
  {Toft} S.,   {Man} A.,  2018, \mn@doi [\apj] {10.3847/2041-8213/aa9fee},
  \href {https://ui.adsabs.harvard.edu/\#abs/2018ApJ...852L...7E} {852, L7}

\bibitem[\protect\citeauthoryear{Enander \& M{\"o}rtsell}{Enander \&
  M{\"o}rtsell}{2013}]{Enander2013}
Enander J.,  M{\"o}rtsell E.,  2013, \mn@doi [Journal of High Energy Physics]
  {10.1007/JHEP10(2013)031}, 2013, 1

\bibitem[\protect\citeauthoryear{{Estrada} et~al.,}{{Estrada}
  et~al.}{2007}]{2007Estrada}
{Estrada} J.,  et~al., 2007, \mn@doi [\apj] {10.1086/512599}, \href
  {http://adsabs.harvard.edu/abs/2007ApJ...660.1176E} {660, 1176}

\bibitem[\protect\citeauthoryear{{Fassnacht}, {Moustakas}, {Casertano},
  {Ferguson}, {Lucas}  \& {Park}}{{Fassnacht}
  et~al.}{2004}]{2004ApJ...600L.155F}
{Fassnacht} C.~D.,  {Moustakas} L.~A.,  {Casertano} S.,  {Ferguson} H.~C.,
  {Lucas} R.~A.,   {Park} Y.,  2004, \mn@doi [\apjl] {10.1086/379004}, \href
  {http://adsabs.harvard.edu/abs/2004ApJ...600L.155F} {600, L155}

\bibitem[\protect\citeauthoryear{Fraga, Barres de Almeida, Bom, Brandt,
  Giommi, Schubert  \& de Albuquerque}{Fraga et~al.}{2021}]{fraga2021}
Fraga B. M.~O.,  Barres de Almeida U.,  Bom C.~R.,  Brandt C.~H.,  Giommi P.,
   Schubert P.,   de Albuquerque M.~P.,  2021, \mn@doi [Monthly Notices of the
  Royal Astronomical Society] {10.1093/mnras/stab1349}, 505, 1268

\bibitem[\protect\citeauthoryear{{Gavazzi}, {Marshall}, {Treu}  \&
  {Sonnenfeld}}{{Gavazzi} et~al.}{2014}]{RINGFINDER}
{Gavazzi} R.,  {Marshall} P.~J.,  {Treu} T.,   {Sonnenfeld} A.,  2014, \mn@doi
  [\apj] {10.1088/0004-637X/785/2/144}, \href
  {http://adsabs.harvard.edu/abs/2014ApJ...785..144G} {785, 144}

\bibitem[\protect\citeauthoryear{Gilman, Birrer, Treu, Keeton  \&
  Nierenberg}{Gilman et~al.}{2018}]{gilman2018probing}
Gilman D.,  Birrer S.,  Treu T.,  Keeton C.~R.,   Nierenberg A.,  2018, Monthly
  Notices of the Royal Astronomical Society, 481, 819

\bibitem[\protect\citeauthoryear{{Gladders}, {Hoekstra}, {Yee}, {Hall}  \&
  {Barrientos}}{{Gladders} et~al.}{2003}]{2003ApJ...593...48G}
{Gladders} M.~D.,  {Hoekstra} H.,  {Yee} H.~K.~C.,  {Hall} P.~B.,
  {Barrientos} L.~F.,  2003, \mn@doi [\apj] {10.1086/376518}, \href
  {http://adsabs.harvard.edu/abs/2003ApJ...593...48G} {593, 48}

\bibitem[\protect\citeauthoryear{Glazebrook, Jacobs, Collett, More  \&
  McCarthy}{Glazebrook et~al.}{2017}]{10.1093/mnras/stx1492}
Glazebrook K.,  Jacobs C.,  Collett T.,  More A.,   McCarthy C.,  2017, \mn@doi
  [Monthly Notices of the Royal Astronomical Society] {10.1093/mnras/stx1492},
  471, 167

\bibitem[\protect\citeauthoryear{Goodfellow, Bengio  \& Courville}{Goodfellow
  et~al.}{2016}]{Goodfellow-et-al-2016}
Goodfellow I.,  Bengio Y.,   Courville A.,  2016, Deep Learning.
MIT Press

\bibitem[\protect\citeauthoryear{Green et~al.,}{Green
  et~al.}{2012}]{green2012wide}
Green J.,  et~al., 2012, arXiv preprint arXiv:1208.4012

\bibitem[\protect\citeauthoryear{{Grillo}, {Rosati}, {Suyu}, {Caminha},
  {Mercurio}  \& {Halkola}}{{Grillo} et~al.}{2020}]{2020ApJ...898...87G}
{Grillo} C.,  {Rosati} P.,  {Suyu} S.~H.,  {Caminha} G.~B.,  {Mercurio} A.,
  {Halkola} A.,  2020, \mn@doi [\apj] {10.3847/1538-4357/ab9a4c}, \href
  {https://ui.adsabs.harvard.edu/abs/2020ApJ...898...87G} {898, 87}

\bibitem[\protect\citeauthoryear{{Guy} et~al.,}{{Guy}
  et~al.}{2022}]{2022zndo...5836022G}
{Guy} L.~P.,  et~al., 2022, in Zenodo id. 5836022. p. 5836022 (\mn@eprint
  {arXiv} {2201.03862}), \mn@doi{10.5281/zenodo.5836022}

\bibitem[\protect\citeauthoryear{Hassan, Islam, Uddin, Ghoshal, Hassan, Huda
  \& Fortino}{Hassan et~al.}{2022}]{cancer_lime}
Hassan M.~R.,  Islam M.~F.,  Uddin M.~Z.,  Ghoshal G.,  Hassan M.~M.,  Huda S.,
    Fortino G.,  2022, \mn@doi [Future Generation Computer Systems]
  {https://doi.org/10.1016/j.future.2021.09.030}, 127, 462

\bibitem[\protect\citeauthoryear{{Haunschmid}, {Manilow}  \&
  {Widmer}}{{Haunschmid} et~al.}{2020}]{SLime}
{Haunschmid} V.,  {Manilow} E.,   {Widmer} G.,  2020, arXiv e-prints, \href
  {https://ui.adsabs.harvard.edu/abs/2020arXiv200902051H} {p. arXiv:2009.02051}

\bibitem[\protect\citeauthoryear{He, Zhang, Ren  \& Sun}{He
  et~al.}{2016}]{he2016deep}
He K.,  Zhang X.,  Ren S.,   Sun J.,  2016, in Proceedings of the IEEE
  conference on computer vision and pattern recognition. pp 770--778

\bibitem[\protect\citeauthoryear{Hezaveh, Dalal, Holder, Kisner, Kuhlen  \&
  Levasseur}{Hezaveh et~al.}{2014}]{hezaveh2014measuring}
Hezaveh Y.,  Dalal N.,  Holder G.,  Kisner T.,  Kuhlen M.,   Levasseur L.~P.,
  2014, arXiv preprint arXiv:1403.2720

\bibitem[\protect\citeauthoryear{{Hezaveh}, {Levasseur}  \&
  {Marshall}}{{Hezaveh} et~al.}{2017}]{2017Natur.548..555H}
{Hezaveh} Y.~D.,  {Levasseur} L.~P.,   {Marshall} P.~J.,  2017, \mn@doi [\nat]
  {10.1038/nature23463}, \href
  {http://adsabs.harvard.edu/abs/2017Natur.548..555H} {548, 555}

\bibitem[\protect\citeauthoryear{{Hogg}, {Blandford}, {Kundic}, {Fassnacht}  \&
  {Malhotra}}{{Hogg} et~al.}{1996}]{1996ApJ...467L..73H}
{Hogg} D.~W.,  {Blandford} R.,  {Kundic} T.,  {Fassnacht} C.~D.,   {Malhotra}
  S.,  1996, \mn@doi [\apjl] {10.1086/310213}, \href
  {http://adsabs.harvard.edu/abs/1996ApJ...467L..73H} {467, L73}

\bibitem[\protect\citeauthoryear{Ivezi\'c et~al.}{Ivezi\'c
  et~al.}{2019}]{ivezic2008lsst}
Ivezi\'c v.,  et~al., 2019, \mn@doi [Astrophys. J.] {10.3847/1538-4357/ab042c},
  873, 111

\bibitem[\protect\citeauthoryear{{Jacobs} et~al.,}{{Jacobs}
  et~al.}{2019}]{2019MNRAS.484.5330J}
{Jacobs} C.,  et~al., 2019, \mn@doi [\mnras] {10.1093/mnras/stz272}, \href
  {https://ui.adsabs.harvard.edu/\#abs/2019MNRAS.484.5330J} {484, 5330}

\bibitem[\protect\citeauthoryear{{Jones}, {Swinbank}, {Ellis}, {Richard}  \&
  {Stark}}{{Jones} et~al.}{2010}]{2010MNRAS.404.1247J}
{Jones} T.~A.,  {Swinbank} A.~M.,  {Ellis} R.~S.,  {Richard} J.,   {Stark}
  D.~P.,  2010, \mn@doi [\mnras] {10.1111/j.1365-2966.2010.16378.x}, \href
  {http://adsabs.harvard.edu/abs/2010MNRAS.404.1247J} {404, 1247}

\bibitem[\protect\citeauthoryear{{Joseph} et~al.,}{{Joseph}
  et~al.}{2014}]{pca_lensfinder}
{Joseph} R.,  et~al., 2014, \mn@doi [\aap] {10.1051/0004-6361/201423365}, \href
  {https://ui.adsabs.harvard.edu/abs/2014A&A...566A..63J} {566, A63}

\bibitem[\protect\citeauthoryear{{Jullo}, {Natarajan}, {Kneib}, {D'Aloisio},
  {Limousin}, {Richard}  \& {Schimd}}{{Jullo}
  et~al.}{2010}]{2010Sci...329..924J}
{Jullo} E.,  {Natarajan} P.,  {Kneib} J.-P.,  {D'Aloisio} A.,  {Limousin} M.,
  {Richard} J.,   {Schimd} C.,  2010, \mn@doi [Science]
  {10.1126/science.1185759}, \href
  {http://adsabs.harvard.edu/abs/2010Sci...329..924J} {329, 924}

\bibitem[\protect\citeauthoryear{{Koopmans}, {Treu}, {Bolton}, {Burles}  \&
  {Moustakas}}{{Koopmans} et~al.}{2006}]{2006ApJ...649..599K}
{Koopmans} L.~V.~E.,  {Treu} T.,  {Bolton} A.~S.,  {Burles} S.,   {Moustakas}
  L.~A.,  2006, \mn@doi [\apj] {10.1086/505696}, \href
  {http://adsabs.harvard.edu/abs/2006ApJ...649..599K} {649, 599}

\bibitem[\protect\citeauthoryear{{Kovner}}{{Kovner}}{1989}]{1989ApJ...337..621K}
{Kovner} I.,  1989, \mn@doi [\apj] {10.1086/167133}, \href
  {http://adsabs.harvard.edu/abs/1989ApJ...337..621K} {337, 621}

\bibitem[\protect\citeauthoryear{{Kubo} \& {Dell'Antonio}}{{Kubo} \&
  {Dell'Antonio}}{2008}]{Kubo2008}
{Kubo} J.~M.,  {Dell'Antonio} I.~P.,  2008, \mn@doi [\mnras]
  {10.1111/j.1365-2966.2008.12880.x}, \href
  {http://adsabs.harvard.edu/abs/2008MNRAS.385..918K} {385, 918}

\bibitem[\protect\citeauthoryear{{Kubo} et~al.,}{{Kubo}
  et~al.}{2010}]{2010ApJ...724L.137K}
{Kubo} J.~M.,  et~al., 2010, \mn@doi [\apjl] {10.1088/2041-8205/724/2/L137},
  \href {http://adsabs.harvard.edu/abs/2010ApJ...724L.137K} {724, L137}

\bibitem[\protect\citeauthoryear{{Lanusse}, {Ma}, {Li}, {Collett}, {Li},
  {Ravanbakhsh}, {Mandelbaum}  \& {P{\'o}czos}}{{Lanusse}
  et~al.}{2018}]{2018MNRAS.473.3895L}
{Lanusse} F.,  {Ma} Q.,  {Li} N.,  {Collett} T.~E.,  {Li} C.-L.,  {Ravanbakhsh}
  S.,  {Mandelbaum} R.,   {P{\'o}czos} B.,  2018, \mn@doi [\mnras]
  {10.1093/mnras/stx1665}, \href
  {https://ui.adsabs.harvard.edu/\#abs/2018MNRAS.473.3895L} {473, 3895}

\bibitem[\protect\citeauthoryear{Laureijs et~al.,}{Laureijs
  et~al.}{2011}]{laureijs2011euclid}
Laureijs R.,  et~al., 2011, arXiv preprint arXiv:1110.3193

\bibitem[\protect\citeauthoryear{{Legin}, {Hezaveh}, {Perreault Levasseur}  \&
  {Wandelt}}{{Legin} et~al.}{2021}]{2021arXiv211205278L}
{Legin} R.,  {Hezaveh} Y.,  {Perreault Levasseur} L.,   {Wandelt} B.,  2021,
  arXiv e-prints, \href {https://ui.adsabs.harvard.edu/abs/2021arXiv211205278L}
  {p. arXiv:2112.05278}

\bibitem[\protect\citeauthoryear{Liu, Jiang, He, Chen, Liu, Gao  \& Han}{Liu
  et~al.}{2020}]{radam}
Liu L.,  Jiang H.,  He P.,  Chen W.,  Liu X.,  Gao J.,   Han J.,  2020, in
  Proceedings of the Eighth International Conference on Learning
  Representations (ICLR 2020).

\bibitem[\protect\citeauthoryear{Magro, Zarb Adami, DeMarco, Riggi  \&
  Sciacca}{Magro et~al.}{2021}]{magro2021}
Magro D.,  Zarb Adami K.,  DeMarco A.,  Riggi S.,   Sciacca E.,  2021, \mn@doi
  [Monthly Notices of the Royal Astronomical Society] {10.1093/mnras/stab1635},
  505, 6155

\bibitem[\protect\citeauthoryear{{Man} et~al.,}{{Man}
  et~al.}{2021}]{2021ApJ...919...20M}
{Man} A. W.~S.,  et~al., 2021, \mn@doi [\apj] {10.3847/1538-4357/ac0ae3}, \href
  {https://ui.adsabs.harvard.edu/abs/2021ApJ...919...20M} {919, 20}

\bibitem[\protect\citeauthoryear{{Marshall} et~al.,}{{Marshall}
  et~al.}{2007}]{Marshal2007}
{Marshall} P.~J.,  et~al., 2007, \mn@doi [\apj] {10.1086/523091}, \href
  {http://adsabs.harvard.edu/abs/2007ApJ...671.1196M} {671, 1196}

\bibitem[\protect\citeauthoryear{{Maturi}, {Mizera}  \& {Seidel}}{{Maturi}
  et~al.}{2014}]{Maturi2014}
{Maturi} M.,  {Mizera} S.,   {Seidel} G.,  2014, \mn@doi [\aap]
  {10.1051/0004-6361/201321634}, \href
  {http://adsabs.harvard.edu/abs/2014A\%26A...567A.111M} {567, A111}

\bibitem[\protect\citeauthoryear{McCully, Keeton, Wong  \& Zabludoff}{McCully
  et~al.}{2017}]{mccully2017quantifying}
McCully C.,  Keeton C.~R.,  Wong K.~C.,   Zabludoff A.~I.,  2017, The
  Astrophysical Journal, 836, 141

\bibitem[\protect\citeauthoryear{{Meneghetti}, {Dolag}, {Tormen}, {Bartelmann},
  {Moscardini}, {Perrotta}  \& {Baccigalupi}}{{Meneghetti}
  et~al.}{2004}]{2004MPLA...19.1083M}
{Meneghetti} M.,  {Dolag} K.,  {Tormen} G.,  {Bartelmann} M.,  {Moscardini} L.,
   {Perrotta} F.,   {Baccigalupi} C.,  2004, \mn@doi [Modern Physics Letters A]
  {10.1142/S0217732304014409}, \href
  {http://adsabs.harvard.edu/abs/2004MPLA...19.1083M} {19, 1083}

\bibitem[\protect\citeauthoryear{{Metcalf}}{{Metcalf}}{2022}]{challenge02}
{Metcalf} R. B. e.~a.,  2022, in preparation

\bibitem[\protect\citeauthoryear{{Metcalf} \& {Petkova}}{{Metcalf} \&
  {Petkova}}{2014}]{2014MNRAS.445.1942M}
{Metcalf} R.~B.,  {Petkova} M.,  2014, \mn@doi [\mnras]
  {10.1093/mnras/stu1859}, \href
  {https://ui.adsabs.harvard.edu/\#abs/2014MNRAS.445.1942M} {445, 1942}

\bibitem[\protect\citeauthoryear{{Metcalf} et~al.,}{{Metcalf}
  et~al.}{2019}]{challenge01}
{Metcalf} R.~B.,  et~al., 2019, \mn@doi [\aap] {10.1051/0004-6361/201832797},
  \href {https://ui.adsabs.harvard.edu/abs/2019A&A...625A.119M} {625, A119}

\bibitem[\protect\citeauthoryear{Mishra, Sturm  \& Dixon}{Mishra
  et~al.}{2017}]{lime_music}
Mishra S.,  Sturm B.~L.,   Dixon S.,  2017, in 18th ISMIR, Suzhou, China. pp
  537 -- 543

\bibitem[\protect\citeauthoryear{{More}, {Cabanac}, {More}, {Alard},
  {Limousin}, {Kneib}, {Gavazzi}  \& {Motta}}{{More} et~al.}{2012}]{2012More}
{More} A.,  {Cabanac} R.,  {More} S.,  {Alard} C.,  {Limousin} M.,  {Kneib}
  J.-P.,  {Gavazzi} R.,   {Motta} V.,  2012, \mn@doi [\apj]
  {10.1088/0004-637X/749/1/38}, \href
  {http://adsabs.harvard.edu/abs/2012ApJ...749...38M} {749, 38}

\bibitem[\protect\citeauthoryear{{More} et~al.,}{{More}
  et~al.}{2016}]{SPACEWARPSII}
{More} A.,  et~al., 2016, \mn@doi [\mnras] {10.1093/mnras/stv1965}, \href
  {http://adsabs.harvard.edu/abs/2016MNRAS.455.1191M} {455, 1191}

\bibitem[\protect\citeauthoryear{{Morgan} et~al.,}{{Morgan}
  et~al.}{2021}]{2021arXiv211201541M}
{Morgan} R.,  et~al., 2021, arXiv e-prints, \href
  {https://ui.adsabs.harvard.edu/abs/2021arXiv211201541M} {p. arXiv:2112.01541}

\bibitem[\protect\citeauthoryear{{Natarajan}, {De Lucia}  \&
  {Springel}}{{Natarajan} et~al.}{2007}]{2007MNRAS.376..180N}
{Natarajan} P.,  {De Lucia} G.,   {Springel} V.,  2007, \mn@doi [\mnras]
  {10.1111/j.1365-2966.2007.11399.x}, \href
  {http://adsabs.harvard.edu/abs/2007MNRAS.376..180N} {376, 180}

\bibitem[\protect\citeauthoryear{{Nord} et~al.,}{{Nord}
  et~al.}{2015}]{DESSL1Nord}
{Nord} B.,  et~al., 2015, arXiv:1512.03062, \href
  {http://adsabs.harvard.edu/abs/2015arXiv151203062N} {}

\bibitem[\protect\citeauthoryear{{Oguri}}{{Oguri}}{2007}]{2007ApJ...660....1O}
{Oguri} M.,  2007, \mn@doi [\apj] {10.1086/513093}, \href
  {http://adsabs.harvard.edu/abs/2007ApJ...660....1O} {660, 1}

\bibitem[\protect\citeauthoryear{Ostrovski et~al.,}{Ostrovski
  et~al.}{2017}]{10.1093/mnrasl/slx173}
Ostrovski F.,  et~al., 2017, \mn@doi [Monthly Notices of the Royal Astronomical
  Society: Letters] {10.1093/mnrasl/slx173}, 473, L116

\bibitem[\protect\citeauthoryear{{Overzier}, {Lemson}, {Angulo}, {Bertin},
  {Blaizot}, {Henriques}, {Marleau}  \& {White}}{{Overzier}
  et~al.}{2013}]{millennium_obs}
{Overzier} R.,  {Lemson} G.,  {Angulo} R.~E.,  {Bertin} E.,  {Blaizot} J.,
  {Henriques} B.~M.~B.,  {Marleau} G.~D.,   {White} S.~D.~M.,  2013, \mn@doi
  [\mnras] {10.1093/mnras/sts076}, \href
  {https://ui.adsabs.harvard.edu/abs/2013MNRAS.428..778O} {428, 778}

\bibitem[\protect\citeauthoryear{{Paraficz} et~al.,}{{Paraficz}
  et~al.}{2016}]{ParaficzCFHTLS}
{Paraficz} D.,  et~al., 2016, arXiv:1605.04309, \href
  {http://adsabs.harvard.edu/abs/2016arXiv160504309P} {}

\bibitem[\protect\citeauthoryear{{Pawase}, {Courbin}, {Faure}, {Kokotanekova}
  \& {Meylan}}{{Pawase} et~al.}{2014}]{HST_galaxyscale_lens}
{Pawase} R.~S.,  {Courbin} F.,  {Faure} C.,  {Kokotanekova} R.,   {Meylan} G.,
  2014, \mn@doi [\mnras] {10.1093/mnras/stu179}, \href
  {https://ui.adsabs.harvard.edu/abs/2014MNRAS.439.3392P} {439, 3392}

\bibitem[\protect\citeauthoryear{{Pearson}, {Maresca}, {Li}  \&
  {Dye}}{{Pearson} et~al.}{2021}]{2021MNRAS.505.4362P}
{Pearson} J.,  {Maresca} J.,  {Li} N.,   {Dye} S.,  2021, \mn@doi [\mnras]
  {10.1093/mnras/stab1547}, \href
  {https://ui.adsabs.harvard.edu/abs/2021MNRAS.505.4362P} {505, 4362}

\bibitem[\protect\citeauthoryear{Petkova, Metcalf  \& Giocoli}{Petkova
  et~al.}{2014}]{petkova2014glamer}
Petkova M.,  Metcalf R.~B.,   Giocoli C.,  2014, Monthly Notices of the Royal
  Astronomical Society, 445, 1954

\bibitem[\protect\citeauthoryear{{Petrillo} et~al.,}{{Petrillo}
  et~al.}{2017}]{2017arXiv170207675P}
{Petrillo} C.~E.,  et~al., 2017, arXiv: 1702.07675, \href
  {http://adsabs.harvard.edu/abs/2017arXiv170207675P} {}

\bibitem[\protect\citeauthoryear{{Petrillo} et~al.,}{{Petrillo}
  et~al.}{2019a}]{2019MNRAS.482..807P}
{Petrillo} C.~E.,  et~al., 2019a, \mn@doi [\mnras] {10.1093/mnras/sty2683},
  \href {https://ui.adsabs.harvard.edu/\#abs/2019MNRAS.482..807P} {482, 807}

\bibitem[\protect\citeauthoryear{{Petrillo} et~al.,}{{Petrillo}
  et~al.}{2019b}]{2019MNRAS.484.3879P}
{Petrillo} C.~E.,  et~al., 2019b, \mn@doi [\mnras] {10.1093/mnras/stz189},
  \href {https://ui.adsabs.harvard.edu/\#abs/2019MNRAS.484.3879P} {484, 3879}

\bibitem[\protect\citeauthoryear{{Pizzuti} et~al.,}{{Pizzuti}
  et~al.}{2016}]{2016arXiv160203385P}
{Pizzuti} L.,  et~al., 2016, arXiv:1602.03385, \href
  {http://adsabs.harvard.edu/abs/2016arXiv160203385P} {}

\bibitem[\protect\citeauthoryear{{Poindexter}, {Morgan}  \&
  {Kochanek}}{{Poindexter} et~al.}{2008}]{2008ApJ...673...34P}
{Poindexter} S.,  {Morgan} N.,   {Kochanek} C.~S.,  2008, \mn@doi [\apj]
  {10.1086/524190}, \href {http://adsabs.harvard.edu/abs/2008ApJ...673...34P}
  {673, 34}

\bibitem[\protect\citeauthoryear{{Ratnatunga}, {Griffiths}  \&
  {Ostrander}}{{Ratnatunga} et~al.}{1999}]{1999AJ....117.2010R}
{Ratnatunga} K.~U.,  {Griffiths} R.~E.,   {Ostrander} E.~J.,  1999, \mn@doi
  [\aj] {10.1086/300840}, \href
  {http://adsabs.harvard.edu/abs/1999AJ....117.2010R} {117, 2010}

\bibitem[\protect\citeauthoryear{Ribeiro, Singh  \& Guestrin}{Ribeiro
  et~al.}{2016}]{lime}
Ribeiro M.~T.,  Singh S.,   Guestrin C.,  2016, in Proceedings of the 22nd
  {ACM} {SIGKDD} International Conference on Knowledge Discovery and Data
  Mining, San Francisco, CA, USA, August 13-17, 2016. pp 1135--1144

\bibitem[\protect\citeauthoryear{{Richard}, {Jones}, {Ellis}, {Stark},
  {Livermore}  \& {Swinbank}}{{Richard} et~al.}{2011}]{2011MNRAS.413..643R}
{Richard} J.,  {Jones} T.,  {Ellis} R.,  {Stark} D.~P.,  {Livermore} R.,
  {Swinbank} M.,  2011, \mn@doi [\mnras] {10.1111/j.1365-2966.2010.18161.x},
  \href {http://adsabs.harvard.edu/abs/2011MNRAS.413..643R} {413, 643}

\bibitem[\protect\citeauthoryear{Russakovsky et~al.,}{Russakovsky
  et~al.}{2015}]{ILSVRC15}
Russakovsky O.,  et~al., 2015, \mn@doi [International Journal of Computer
  Vision (IJCV)] {10.1007/s11263-015-0816-y}, 115, 211

\bibitem[\protect\citeauthoryear{Schuldt, Suyu, Meinhardt, Leal-Taix{\'e},
  Ca{\~n}ameras, Taubenberger  \& Halkola}{Schuldt
  et~al.}{2021}]{schuldt2021holismokes}
Schuldt S.,  Suyu S.,  Meinhardt T.,  Leal-Taix{\'e} L.,  Ca{\~n}ameras R.,
  Taubenberger S.,   Halkola A.,  2021, Astronomy \& Astrophysics, 646, A126

\bibitem[\protect\citeauthoryear{{Schwab}, {Bolton}  \& {Rappaport}}{{Schwab}
  et~al.}{2010}]{Schwab2010}
{Schwab} J.,  {Bolton} A.~S.,   {Rappaport} S.~A.,  2010, \mn@doi [\apj]
  {10.1088/0004-637X/708/1/750}, \href
  {http://adsabs.harvard.edu/abs/2010ApJ...708..750S} {708, 750}

\bibitem[\protect\citeauthoryear{Spiniello et~al.,}{Spiniello
  et~al.}{2018}]{10.1093/mnras/sty1923}
Spiniello C.,  et~al., 2018, \mn@doi [Monthly Notices of the Royal Astronomical
  Society] {10.1093/mnras/sty1923}, 480, 1163

\bibitem[\protect\citeauthoryear{{Suyu}, {Marshall}, {Auger}, {Hilbert},
  {Blandford}, {Koopmans}, {Fassnacht}  \& {Treu}}{{Suyu}
  et~al.}{2010}]{2010ApJ...711..201S}
{Suyu} S.~H.,  {Marshall} P.~J.,  {Auger} M.~W.,  {Hilbert} S.,  {Blandford}
  R.~D.,  {Koopmans} L.~V.~E.,  {Fassnacht} C.~D.,   {Treu} T.,  2010, \mn@doi
  [\apj] {10.1088/0004-637X/711/1/201}, \href
  {http://adsabs.harvard.edu/abs/2010ApJ...711..201S} {711, 201}

\bibitem[\protect\citeauthoryear{Tan \& Le}{Tan \& Le}{2019}]{efficientnet}
Tan M.,  Le Q.,  2019, in International Conference on Machine Learning. pp
  6105--6114

\bibitem[\protect\citeauthoryear{Tan, Chen, Pang, Vasudevan, Sandler, Howard
  \& Le}{Tan et~al.}{2019}]{tan2019mnasnet}
Tan M.,  Chen B.,  Pang R.,  Vasudevan V.,  Sandler M.,  Howard A.,   Le Q.~V.,
   2019, in Proceedings of the IEEE Conference on Computer Vision and Pattern
  Recognition. pp 2820--2828

\bibitem[\protect\citeauthoryear{{Treu} \& {Koopmans}}{{Treu} \&
  {Koopmans}}{2002a}]{2002MNRAS.337L...6T}
{Treu} T.,  {Koopmans} L.~V.~E.,  2002a, \mn@doi [\mnras]
  {10.1046/j.1365-8711.2002.06107.x}, \href
  {http://adsabs.harvard.edu/abs/2002MNRAS.337L...6T} {337, L6}

\bibitem[\protect\citeauthoryear{Treu \& Koopmans}{Treu \&
  Koopmans}{2002b}]{0004-637X-575-1-87}
Treu T.,  Koopmans L. V.~E.,  2002b, The Astrophysical Journal, 575, 87

\bibitem[\protect\citeauthoryear{{Vegetti}, {Lagattuta}, {McKean}, {Auger},
  {Fassnacht}  \& {Koopmans}}{{Vegetti} et~al.}{2012}]{2012Natur.481..341V}
{Vegetti} S.,  {Lagattuta} D.~J.,  {McKean} J.~P.,  {Auger} M.~W.,  {Fassnacht}
  C.~D.,   {Koopmans} L.~V.~E.,  2012, \mn@doi [\nat] {10.1038/nature10669},
  \href {https://ui.adsabs.harvard.edu/\#abs/2012Natur.481..341V} {481, 341}

\bibitem[\protect\citeauthoryear{{Walmsley} et~al.,}{{Walmsley}
  et~al.}{2022}]{2022MNRAS.509.3966W}
{Walmsley} M.,  et~al., 2022, \mn@doi [\mnras] {10.1093/mnras/stab2093}, \href
  {https://ui.adsabs.harvard.edu/abs/2022MNRAS.509.3966W} {509, 3966}

\bibitem[\protect\citeauthoryear{{Wen}, {Han}  \& {Jiang}}{{Wen}
  et~al.}{2011}]{2011RAA....11.1185W}
{Wen} Z.-L.,  {Han} J.-L.,   {Jiang} Y.-Y.,  2011, \mn@doi [Research in
  Astronomy and Astrophysics] {10.1088/1674-4527/11/10/007}, \href
  {http://adsabs.harvard.edu/abs/2011RAA....11.1185W} {11, 1185}

\bibitem[\protect\citeauthoryear{{Wilde}, {Serjeant}, {Bromley}, {Dickinson},
  {Koopmans}  \& {Metcalf}}{{Wilde} et~al.}{2022}]{2022arXiv220212776W}
{Wilde} J.,  {Serjeant} S.,  {Bromley} J.~M.,  {Dickinson} H.,  {Koopmans} L.
  V.~E.,   {Metcalf} R.~B.,  2022, arXiv e-prints, \href
  {https://ui.adsabs.harvard.edu/abs/2022arXiv220212776W} {p. arXiv:2202.12776}

\bibitem[\protect\citeauthoryear{Wong et~al.}{Wong et~al.}{2020}]{Wong:2019kwg}
Wong K.~C.,  et~al., 2020, \mn@doi [Mon. Not. Roy. Astron. Soc.]
  {10.1093/mnras/stz3094}, 498, 1420

\bibitem[\protect\citeauthoryear{{Yamamoto}, {Kadoya}, {Murata}  \&
  {Futamase}}{{Yamamoto} et~al.}{2001}]{2001PThPh.106..917Y}
{Yamamoto} K.,  {Kadoya} Y.,  {Murata} T.,   {Futamase} T.,  2001, \mn@doi
  [Progress of Theoretical Physics] {10.1143/PTP.106.917}, \href
  {http://adsabs.harvard.edu/abs/2001PThPh.106..917Y} {106, 917}

\bibitem[\protect\citeauthoryear{{Zackrisson} \& {Riehm}}{{Zackrisson} \&
  {Riehm}}{2010}]{2010AdAst2010E...9Z}
{Zackrisson} E.,  {Riehm} T.,  2010, \mn@doi [Advances in Astronomy]
  {10.1155/2010/478910}, \href
  {http://adsabs.harvard.edu/abs/2010AdAst2010E...9Z} {2010}

\makeatother
\end{thebibliography}

\end{document}